\documentclass[a4paper,11pt,preprintnumbers]{article}
\pdfoutput=1 % if your are submitting a pdflatex (i.e. if you have
\usepackage{jheppub} % for details on the use of the package, please
\usepackage[T1]{fontenc} % if needed
\usepackage{lmodern,bbm}
\usepackage{graphicx}
\usepackage{xspace}
\usepackage{color}
\usepackage{multirow}
\usepackage[small]{subfigure}
\usepackage{marginnote}
\usepackage{slashed}
\usepackage{ulem}

\newcommand{\nn}{\nonumber} 
\newcommand{\bn}{{\bar n}}
\newcommand{\nz}{n_z}
\newcommand{\bnz}{\bar{n}_z}
\newcommand{\nJ}{n_J}

\newcommand{\mcdot}{\!\cdot\!}

\newcommand{\be}{\begin{equation}}
\newcommand{\ee}{\end{equation}}
\newcommand{\bes}{\begin{subequations} \begin{align} }
\newcommand{\ees}{\end{subequations}\end{align} }
\newcommand{\bea}{\begin{eqnarray}}
\newcommand{\eea}{\end{eqnarray}}

\newcommand{\EICC}{{\textrm {EicC} }}

\newcommand{\bra}[1]{\left\langle #1\right\rvert}
\newcommand{\ket}[1]{\left\lvert #1\right\rangle}

\newcommand{\Lqcd}{\Lambda_{\text{QCD}}}

\newcommand{\as}{\alpha_s}
\newcommand{\MSbar}{\overline{\text{MS}}}

\newcommand{\Ga}{\Gamma}

\newcommand{\cO}{\mathcal{O}}
\newcommand{\cM}{\mathcal{M}}

\newcommand{\cL}{\mathcal{L}}
\newcommand{\cA}{\mathcal{A}}

\newcommand{\cB}{\mathcal{B}}

\newcommand{\e}{\epsilon}

\newcommand{\tW}{{\widetilde W }}
\newcommand{\tR}{{\widetilde R }}
\newcommand{\tN}{{\widetilde N }}

\newcommand{\eq}[1]{Eq.~\eqref{eq:#1}}
\newcommand{\eqs}[2]{Eqs.~\eqref{eq:#1} and \eqref{eq:#2}}
\newcommand{\eqss}[3]{Eqs.~\eqref{eq:#1}, \eqref{eq:#2}, and \eqref{eq:#3}}
\renewcommand{\sec}[1]{Sec.~\ref{sec:#1}}

\newcommand{\appx}[1]{App.~\ref{app:#1}}
\newcommand{\fig}[1]{Fig.~\ref{fig:#1}}

\newcommand{\taub}{\tau^b}
\newcommand{\taujt}{\tau^{\text{jt}}}
\newcommand{\minnJ}{\mathop{\textrm{min}}_{\hat{n}_{J}}}
\newcommand{\taukt}{\tau^\text{kt}}
\newcommand{\tauct}{\tau^\text{ct}}
\newcommand{\taujtkt}{\tau^\text{jt,kt}}
\newcommand{\zct}{z_\text{ct}}
\newcommand{\zjt}{z_\text{jt}}

\newcommand{\comment}[1]{}

\newcommand{\sing}{\text{sing} }
\newcommand{\ns}{\text{ns} }
 
\newcommand{\cumulant}{{\rm c}}
\newcommand{\sigmac}{\sigma^\cumulant}

\newcommand{\onet}{\frac{1}{1+\tau}}
\newcommand{\onez}{\frac{1-z}{z}}

\newcommand{\deltatz}{\theta\left(z-\tfrac{2}{3}\right)\delta\left(\tau-\tfrac{1-z}{z}\right)}
\newcommand{\deltaz}{\delta\left(\tau-\tfrac{1-z}{z}\right)}
\newcommand{\taum}{\tau_{\text{m}}}

\newcommand{\Ans}{A^\text{ns}}
\newcommand{\Bns}{B^\text{ns}}
\newcommand{\cAns}{\cA^\text{ns}}
\newcommand{\cBns}{\cB^\text{ns}}
\newcommand{\dAns}{ \delta A^\text{ns}}
\newcommand{\dBns}{ \delta B^\text{ns}}
\newcommand{\cns}{c_\text{ns}} 

\newcommand{\intmax}{ \int_{\max\big[ \frac{1}{1+\tau},\frac{2}{3},x \big]}^1}
\newcommand{\intd}{\int_{\zct}^{\zjt}}
\newcommand{\Thfour}{{\Theta_\text{IV}}}

\allowdisplaybreaks[1]

\newcommand{\Pqqr}{\frac{1+z^2}{1-z}}

\newcommand{\ie}{\textit{i}.\textit{e}., }

\begin{document}

\today
\title{1-jettiness with jet axis at $O(\alpha_s)$
in deep inelastic scattering}

\author[*]{Zexuan Chu,}
\author[*]{Yunlu Wang,}
\author[*]{June-Haak Ee,}
\author[]{Jinhui Chen,}
\author[]{Daekyoung Kang}
\note[*]{These authors made main contributions equally to this work.}
\affiliation[]{Key Laboratory of Nuclear Physics and Ion-beam Application (MOE) and Institute of Modern Physics,
Fudan University, \\
Shanghai 200433, China }

%% e-mail addresses: one for each author, in the same order as the authors

\emailAdd{zxchu19@fudan.edu.cn}
\emailAdd{yunluwang20@fudan.edu.cn}
\emailAdd{june\_haak\_ee@fudan.edu.cn}
\emailAdd{chenjinhui@fudan.edu.cn}
\emailAdd{dkang@fudan.edu.cn}

%%%%%%%%%%%%%%%%%%%%%%%%%%%%%%%%%%%%%%%%%%%%%%%%%%%%%%
\abstract{
We present $O(\alpha_s)$ analytic predictions for event shape 1-jettiness $\tau_1$ distribution aiming measurements in deep inelastic scattering process at future Electron Ion Colliders. 
The result depends on conventional variables $x$ and $Q$ as well as on $\tau_1$ and is relatively compact and easy to implement for numerical calculation. Three different choices of axis, with respect to which $\tau_1$ is measured are considered in the Breit frame.
The first is the one optimally adjusted to minimize $\tau_1$ and the second and third are taken from anti-$k_T$ and Centauro jet algorithms defined with a jet radius parameter $R$, respectively.
We find that the first and second give the same result at this order and are independent of $R$,
while the third depends on the radius. This fixed-order result provides a nonsingular contribution to be combined with a singular log-resummed contribution to give the full spectrum in $\tau_1$ space and also shows how fixed-order and resummation regions change as a function of $x$ and $Q$.}
%%%%%%%%%%%%%%%%%%%%%%%%%%%%%%%%%%%%%%%%%%%%%%%%%%%%%%

\maketitle

%\vspace*{40cm}
%\clearpage

%%%%%%%%%%%%%%%%%%%%%%%%%%%%%%%%%%%%%%%%%%%%%%%%%%%%%%
\section{Introduction}\label{sec:intro}
%%%%%%%%%%%%%%%%%%%%%%%%%%%%%%%%%%%%%%%%%%%%%%%%%%%%%%
Jets, energetic hadron bunches produced in high-energy collisions,  
are useful tool to study the strong dynamics induced by Quantum Chromodynamics (QCD) and to probe 
new physics beyond the Standard Model. Its application to future QCD machines that probe internal structure of ions with electron beam is extensively investigated in Electron-Ion-Collider in the US (EIC) \cite{AbdulKhalek:2021gbh,Accardi:2012qut} and in Electron-ion-collider in China ($\EICC$) \cite{Anderle:2021wcy}.
The jets can be defined and studied exclusively using jet finding algorithms \cite{Catani:1991hj,Catani:1993hr,Ellis:1993tq,Dokshitzer:1997in,Salam:2007xv,Cacciari:2008gp},
which conventionally require parameters like a jet radius and jet veto.
On the other hand, they can also be studied inclusively with classic observables called event shapes \cite{Dasgupta:2003iq},
which are theoretically simpler with a small number of parameters, hence easier to achieve higher accuracy compared to exclusive jet study.  An example of event shapes is a thrust which has been predicted up to N$^3$LL$+\cO(\as^3)$ accuracy \cite{GehrmannDeRidder:2007bj,GehrmannDeRidder:2007hr,Weinzierl:2008iv,Weinzierl:2009ms,Becher:2008cf,Abbate:2010xh}  in $e^+e^-$ annihilation. From the event shape, the strong coupling constant was determined at 1\% precision \cite{Becher:2008cf,Chien:2010kc,Abbate:2010xh}, which is one among the most precise determinations listed in Particle Data Book \cite{ParticleDataGroup:2020ssz}. 

The thrust as well as other event shapes in deep-inelastic scattering (DIS) was studied \cite{Dasgupta:2003iq} and measured in HERA experiment \cite{Adloff:1997gq,Adloff:1999gn,Aktas:2005tz,Breitweg:1997ug,Chekanov:2002xk,Chekanov:2006hv}. 
Due to limited detector coverage they were defined from particles in a current hemisphere, to which products of hard scattering usually belong while initial-state radiations and beam remnants do not. On the other hand, the definition in $e^+e^-$ is done with both hemispheres. 
More recent developments including improved accuracy and/or new event shape predictions \cite{Kang:2013nha,Kang:2013lga,Aschenauer:2019uex,Li:2020bub,Kang:2020fka,Li:2021txc,Zhu:2021xjn}
assume that future machines can cover both regions. 
With the assumption, DIS thrust, which we call a 1-jettiness \cite{Kang:2013nha,Kang:2013lga} can be defined as
\footnote{It is called 1-jettiness since it is  $N=1$ version of N-jettiness \cite{Stewart:2010tn}
that is generalized thrust for N-jet events} 
%%%
\be \label{eq:tau}
\tau_1 = \frac{2}{Q^2}\sum_{i\in X}\min\{ q_B\mcdot p_i,q_J\mcdot p_i\}
 \,,
\ee
%%%
where $p_i$ is the momentum of particle $i$ in the final state $X$ and $Q$ is a hard momentum transferred
by a virtual photon. $q_B$ and $q_J$ are beam and jet axes, onto which momentum $p_i$ is projected. The operator `min' takes smaller one among two scalar products and this makes particles grouped into one of two regions, beam or jet.  The value of $\tau_1$ is small when the final state $X$ contains two collimated bunches along each
of beam and jet axes. Otherwise, for multi-jet final state the value is not small.
In \cite{Kang:2013nha}, a version called $\tau_1^b$ was defined by $z$ axis in the Breit frame for $q_J$ and known to be same as a version of DIS thrust called $\tau_Q$ measured in HERA \cite{Antonelli:1999kx}. 
It was computed analytically at the first order in $\alpha_s$ \cite{Kang:2014qba}.

In this paper, we study another version called $\tau_1^a$ \cite{Kang:2012zr,Kang:2013nha} for which the jet axis $q_J$ is determined by axis finding algorithms, while $q_B$ defined next section is held fixed to the beam axis. Its distribution was numerically computed in the Lab frame \cite{Kang:2013lga} by using the axis obtained from anti-$k_T$ algorithms. In this paper we consider three different algorithms in the Breit frame in \sec{1jettiness} and show analytic expressions of our fixed-order predictions at the first order in $\alpha_s$ and separately next-to-leading power (NLP) terms in small $\tau_1$ limit in \sec{xsec} as well as numerical results in \sec{num}. Finally, we summarize in \sec{con}.

%%%%%%%%%%%%%%%%%%%%%%%%%%%%%%%%%%%%%%%%%%%%%%%%%%%%%%
\section{1-jettiness with three jet axes} \label{sec:1jettiness}
%%%%%%%%%%%%%%%%%%%%%%%%%%%%%%%%%%%%%%%%%%%%%%%%%%%%%%
In this section we define the jet axis $q_J$ and write down the expression of 1-jettiness defined with the axis for 2-body final state.
We consider three different jet axes in the Breit frame. The first axis is the axis that minimizes the value of 1-jettiness  as in $e^+e^-$ thrust and we call it a 1-jettiness axis. Next two axes are obtained from jet momentum defined by anti-$k_T$ algorithm \cite{Cacciari:2008gp} and by Centauro algorithm \cite{Arratia:2020ssx}. 

We first make a quick review of kinematics in the Breit frame.
In the frame, the virtual photon with momentum $q$ and the proton with
momentum $P$ are aligned along the $z$ axis.  One of advantages of this frame is that
the initial state radiation moving along the proton direction is well separated from the other particles produced by a hard scattering.
We take the proton to move against $z$ direction
%%%
\be\label{eq:Pmu}
P^\mu = \frac{Q}{x}\frac{\bnz^\mu}{2}\,,
\ee
%%%
where the unit vectors $\nz= (1,0,0,1)$ and $\bnz=(1,0,0,-1)$, the hard scale $Q$ is the photon virtuality $Q^2=-q^2$ and the Bj\"orken variable $x=Q^2/(2P\mcdot q)$.
The virtual photon $q$ is spacelike
%%%
\be\label{eq:qmu}
q^\mu = Q\frac{\nz^\mu-\bnz^\mu}{2}\,.
\ee
%%%

A parton from the proton taking a fractional momentum $p=\xi P=Q/z \tfrac\bnz2$, where $z=x/\xi$
and $x<z<1$, scatters off the photon.
At order $\cO(\as)$, the final states contain two particles with momenta $p_1$ and $p_2$.
Using the momentum conservation $p+q=p_1+p_2$ and the onshell condition $p_1^2=p_2^2=0$ 
we can express the final momenta in terms of $z$ and a dimensionless variable $v= p^-_2 / Q$ as
%%%
\begin{align}\label{eq:p1p2}
 p_1^\mu &= Q(1-v)\frac{\nz^\mu}{2}+Q\frac{1-z}{z}v\frac{\bnz^\mu}{2}
 -Q\sqrt{\frac{1-z}{z}(1-v)v}~n_\perp^\mu,
 \\ \nn
 p_2^\mu &= Q v\frac{\nz^\mu}{2}+Q\frac{1-z}{z}(1-v)\frac{\bnz^\mu}{2}
 +Q\sqrt{\frac{1-z}{z}(1-v)v}~n_\perp^\mu,
\end{align}
%%%
where $0<v<1$ and $n_\perp$ is an orthogonal unit vector to $\nz$ and $\bnz$, \ie $n_\perp\mcdot \nz=n_\perp\mcdot \bnz=0$ and $n_\perp\cdot n_\perp=-1$.

The beam axis $q_B$ is defined to be proportional to the proton momentum
%%%
\be\label{eq:qB}
q_B^\mu= x P^\mu= Q\frac{\bnz^\mu}{2}
\,.
\ee
%%%
The scalar products of the beam axis $q_B$ and $p_i$ contributing to 1-jettiness can be expressed as
%%%
\begin{align}\label{eq:qBp12}
& 2q_B\mcdot p_1 = (1-v) Q^2
\,, \nn \\
& 2q_B\mcdot p_2 = v Q^2
\,,\nn\\
& 2q_B\mcdot (p_1+p_2) = Q^2
\,.
\end{align}
%%%

%%%%%%%%%%%%%%%%%%%%%%%%%%%%%%%%%%%%%%%%%%%%%%%%%%%%%%
\subsection{1-jettiness axis $\taujt$} \label{ssec:jt}
\begin{figure}[t]{
\begin{center}
\vspace{-1em}   
\includegraphics[scale=.5]{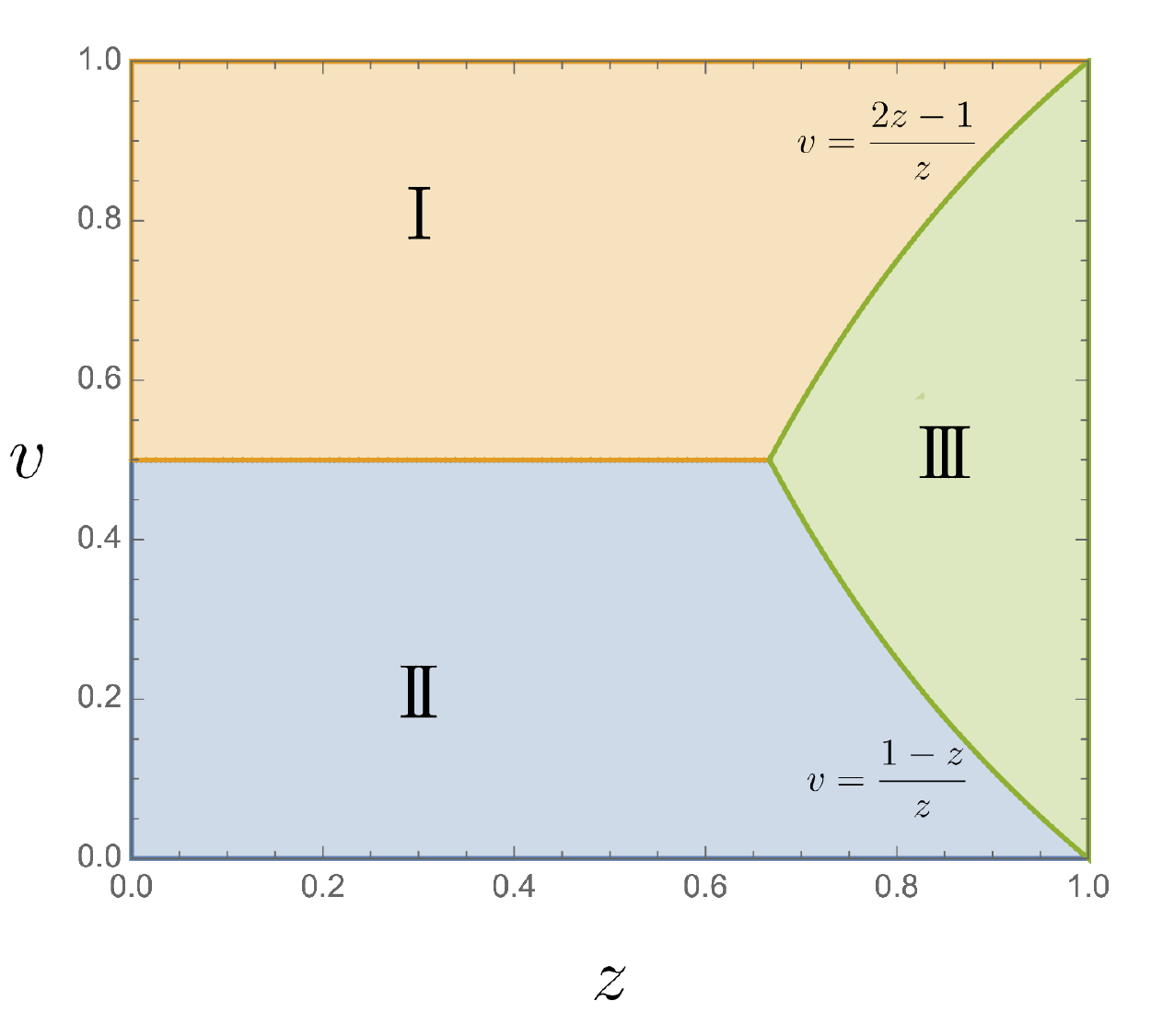}
\includegraphics[scale=.5]{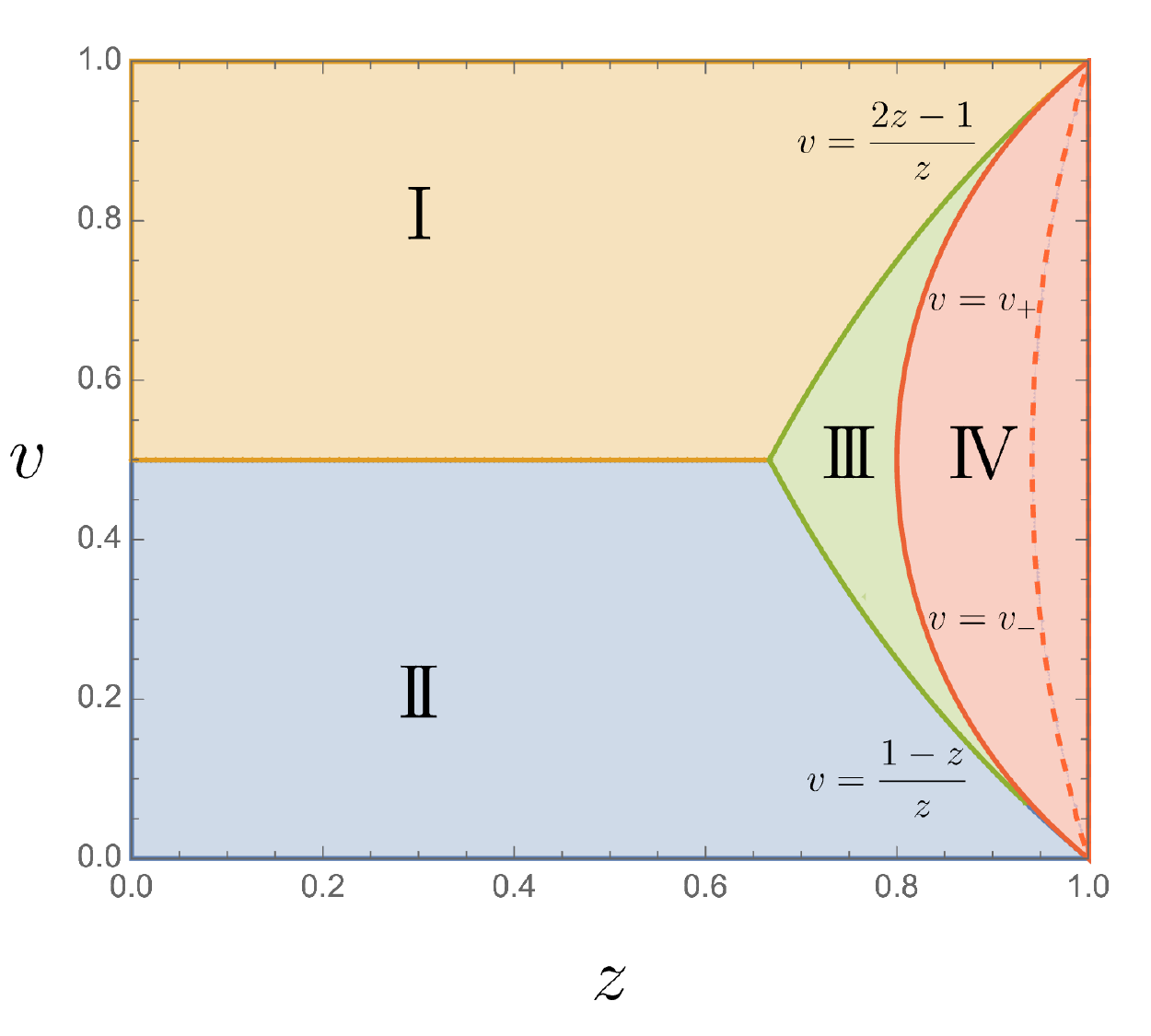}
     \vspace{-2em} 
\end{center}
    { \caption[1]{ 
Two-body phase space for 1-jettiness, $\taujtkt$ with jettiness/anti-$k_T$ axis (left) and $\tauct$ with Centauro axis (right). 1-jettiness takes the same  expression in first three regions while the fourth region is only for $\tauct$. The bound of fourth region for jet radius $R=1$ and $2$ is shown in red dashed and red solid, respectively. 
  \label{fig:vz-space}} }
  \vspace{-1em}
  }
\end{figure}

%%%%%%%%%%%%%%%%%%%%%%%%%%%%%%%%%%%%%%%%%%%%%%%%%%%%%%
In the determination of 1-jettiness axis, we allow the direction of jet axis to be varied, while the absolute magnitude of three-momentum part to be held fixed to $Q/2$. Then, the $q_J$ can be expressed as  
%%%
\be \label{eq:qJ}
q_J^\mu=Q\frac{\nJ^\mu}{2}= \frac{Q}{2}(1,\hat n_J)^\mu
\,,
\ee
%%%
where $\hat n_J$ is a unit vector adjusted event by event in such a way that minimizes the value of 1-jettiness.
Rewriting \eq{tau} with $p_1$ and $p_2$ explicitly gives\footnote{From now on we drop the subscript ``1'' in 1-jettiness $\tau_1$ for a conventional simplicity hence, $\taujt$ implies $\tau_1^{jt}$ and same for $\taub$}

%%%
\begin{align}\label{eq:taujt}
\taujt
&=\frac{2}{Q^2} \minnJ
\left\{ \sum^2_{i=1}
\textrm{min}\{ q_B \mcdot p_i ,q_J \mcdot p_i\}
\right\}
\nn \\
&= \text{min}\left\{
 q_B \mcdot (p_1+p_2),\;
 \minnJ \{q_J \mcdot (p_1+p_2)\},\;
 q_B\mcdot p_1+\minnJ \{ q_J\mcdot p_2\},\;
\minnJ \{ q_J\mcdot p_1 \}+q_B\mcdot p_2 
\right\}
\,\nn \\
&= \textrm{min}
\left\{\; 1  \,,\; \theta \left(-z+\frac{1}{2}\right)+\frac{1-z}{z}\theta\left({z-\frac{1}{2}}\right)\,
,\;1-v \,,\; v
\right\}
\nn \\
&= (1-v)~ \Theta_\text{I}(v,z)
+v~\Theta_\text{II}(v,z)
 +\frac{1-z}{z}~\Theta_\text{III}(v,z) \,.
\end{align}
%%%
The outer `min' operator with $\hat n_J$ beneath it in the first line implies that it is adjusted to minimize quantities in the braces.
In the second line, we listed all four combinations of scalar products between $q_{B,J}$ and $p_{1,2}$ then, moved the outer `min' inside so that the axis $\hat n_J$ is determined by minimizing each scalar product. 
For the first product there is no product with $q_J$ and $\hat n_J$ is not determined.
From the second to the last $\hat n_J=\hat z,\hat p_2,$ and $\hat p_1$, respectively. We used \eq{qBp12} to express the scalar products in terms of $v$ and $z$ and obtained the third line. Then, the outer `min' selects the smallest among those four and it leads to three regions denoted by $\Theta_i$
%%%
\begin{align}\label{eq:Theta}
\Theta_\text{I} &= \theta\left(v-\frac{1}{2}\right)\theta\left(v-\frac{2z-1}{z}\right)
\,,\nn\\
\Theta_\text{II} &= \theta\left(-v+\frac{1}{2}\right)
\theta\left(-v+\frac{1-z}{z}\right)
\,,\nn\\
\Theta_\text{III} &= 
\theta\left(-v+\frac{2z-1}{z}\right)
\theta\left(v-\frac{1-z}{z}\right)
\,,
\end{align}
%%%
where the step function $\theta(x)$ is 1 for $x>0$ and 0 otherwise.
The three regions are plotted in the left panel of \fig{vz-space}. 
From the expression we find the maximum value of $\taujt$
%%%
\be \label{eq:taumax}
\taujt_{\max}=\frac{1}{2}\theta\left(-x+\frac23\right)+\frac{1-x}{x}\theta\left(x-\frac23\right),
\ee
%%%
which is a constant in $v$. We used the hadronic variable $x$  in advance instead of the partonic variable $z$,
which is replaced by $x$ at the end. The maximum occurs on the heavy lines in \fig{vz-space}.
%%%
%%%

%%%%%%%%%%%%%%%%%%%%%%%%%%%%%%%%%%%%%%%%%%%%%%%%%%%%%%
\subsection{Anti-$k_T$ algorithm $\taukt$} 
%%%%%%%%%%%%%%%%%%%%%%%%%%%%%%%%%%%%%%%%%%%%%%%%%%%%%%
In the $k_T$-type algorithms \cite{Cacciari:2008gp} designed to be invariant under longitudinal boosts,
the distance between particles is defined in term of transverse momentum $\mathbf{p}_{T}$ and invariant angular distances
%%%
\begin{align}\label{eq:dij}
d_{ij}=d_{ji}&=\min\big\{\mathbf{p}_{Ti}^{2p} \,, \mathbf{p}_{Tj}^{2p} \big\}\frac{\Delta R^2_{ij}}{R^2},\quad d_{iB}=\mathbf{p}_{Ti}^{2p}\,,
 \\
 \label{eq:DRij}
\Delta R^2_{ij}&=(y_i-y_j)^2 + (\varphi_i-\varphi_j)^2\,,
\end{align}
%%%
where $d_{ij}$ is the  distance between outgoing particles $i$ and $j$, and $d_{iB}$ is the distance of a particle $i$ from the beam.
$\mathbf{p}_{Ti}$ is the magnitude of the transverse momentum 
of particle $i$, $R$ is a jet radius parameter of $\cO(1)$ and $\Delta R_{ij}$ is an angular distance defined in terms of distances in rapidity $y$ and in azimuth $\varphi$.
 The power $p$ on the momentum is a real valued parameter. When $p= 1,0, -1$, it is called $k_T$, Cambridge/Aachen and anti-$k_T$, respectively.  In this paper we take the anti-$k_T$ algorithm $p=-1$, which is a conventional choice by many experiments.

With above distances, the jet algorithm finds minimum $d_\text{min}$ among all the $d_{ij}$ and $d_{iB}$. If $d_\text{min}$ is a $d_{ij}$, particle $i$ and $j$ are merged into a single particle and if $d_\text{min}$ is a $d_{iB}$, the particle is declared to be a final jet and removed from the list of final particles. We repeat this with new list of particles until all the particles in the list are gone.

For a two-body final state, we need to compare $d_{12},d_{1B}$ and $d_{2B}$ to classify jets. Recall that in the Breit frame the transverse momenta of two particles in \eq{p1p2} are back-to-back so that
%%%
\be \label{eq:d1B}
d_{1B}=d_{2B}=\vert{\mathbf{p}_{T1}}\vert^{-2}
\,.
\ee
%%%
The distance in azimuthal angle between $p_{1,2}$ is $(\varphi_1-\varphi_2)^2 =\pi^2$ and the distance in rapidity is $ (y_1-y_2)^2=\ln^2\tfrac{1-v}{v}$. Then, the distance $d_{ij}$ is
%%%
\be \label{eq:d12}
d_{12} =\vert{\mathbf{p}_{T1}}\vert^{-2} \frac{\pi^2+\ln^2\frac{1-v}{v}}{R^2}
\,.
\ee
%%%
Typical size of jet radius $R$ is smaller than $\pi$ and $\ln^2\tfrac{1-v}{v}>0$ hence, $d_{ij}$ is always greater than $d_{1B}$. This means that there is no merging in the two-body final state. The anti-$k_T$ algorithm gives two jets and each of them is $p_1$ and $p_2$ and the jet momentum of each jet is $\mathbf{p}_i$. The jet axis $q_J$ is defined from the jet momentum as 
%%%
\be\label{eq:qjkt}
q_{J}^{kt}=\big\{  |\mathbf{p}_1|(1,\hat{\mathbf p}_1), |\mathbf{p}_2|(1,\hat{\mathbf p}_2)\big\}= \big\{p_1, p_2 \big\} \,,
\ee
%%%
where in the second equality, we used the fact that $p_i$ is massless. In higher multiplicity, the jet algorithm returns more jets. Among those jets, what we need is collinear jets with a large momentum that are candidates for the jet axis $q_J$ and by implementing a veto condition we can reject unnecessary soft jets with a small momentum. Among the candidates, the one that gives the smallest value of 1-jettiness will be selected.
Then, 1-jettiness with the candidates $p_1$ and $p_2$ is given by
%%%
\begin{align} \label{eq:tau-kt-2body}
  \taukt &=
   \frac{2}{Q^2}\min_{ q_J\in \left\{p_1,p_2\right\}}
   \bigg\{\sum_{i=1}^2 \min \{q_B\mcdot p_i\,,q_J\mcdot p_i \} \bigg\}
   \\ \nn
  &= \frac{2}{Q^2}
  \min\bigg\{
  \sum_{i=1}^2 \min \{q_B\mcdot p_i\,,q_1\mcdot p_i \}\,,\,
  \sum_{i=1}^2 \min\{q_B\mcdot p_i\,,q_2\mcdot p_i \} 
  \bigg\}
  \\ \nn
  &=\frac{2}{Q^2}\textrm{min}\bigg\{
  \min\{q_B\mcdot (p_1+p_2),\,p_1\mcdot(p_1+p_2),\,q_B\mcdot p_1+p_1\mcdot p_2\,,\,q_B\mcdot p_2+p_1\mcdot p_1\}\,,\,\nn\\
  &\qquad\qquad\qquad \min\{q_B\mcdot (p_1+p_2),\,p_2\mcdot (p_1+p_2),\,q_B\mcdot p_2+p_1\mcdot p_2\,,\,q_B\mcdot p_1+p_2\mcdot p_2\}
  \bigg\}
  \nn\\
  &=\min
  \bigg\{  
  \frac{1-z}{z},\,v,\,1-v
  \bigg\}
  \nn\\
&=\taujt
\,,\nn
\end{align}
%%%
where in the second line we set $q_J=p_1$ and $q_J=p_2$ after first and second summation symbols, respectively.
Then, as was done in \eq{taujt}, all possible combinations are taken into account in the third line.
To get the fifth line, we used \eq{qBp12} then, deleted trivial terms that cannot be taken as a minimum.
Interestingly, we come up with a fact that $\taukt$ is equal to $\taujt$ in the two-body case. This is because of simplification with a small number of final particles. In general, the radius $R$ dependence enters in the case of anti-$k_T$ algorithm and we do not expect that $\taujt$ and $\taukt$ are the same with higher multiplicity.
We do not consider higher multiplicity in this paper and from now on, we do not separately treat $\taukt$ except for the case when we need to distinguish $\taukt$ from $\taujt$. 

%%%%%%%%%%%%%%%%%%%%%%%%%%%%%%%%%%%%%%%%%%%%%%%%%%%%%%
\subsection{Centauro algorithm $\tauct$} 
%%%%%%%%%%%%%%%%%%%%%%%%%%%%%%%%%%%%%%%%%%%%%%%%%%%%%%
For the third jet axis we would like to adopt the Centauro algorithm \cite{Arratia:2020ssx}, which is more recently introduced algorithm for DIS taking into account the target-current asymmetry in the Breit frame.
It is still longitudinal-boost invariant like the anti-$k_T$ algorithm but it allows to capture jets close to beam axis while the anti-$k_T$ algorithm cannot form a jet in the region since the rapidity distances between particles across the beam line become very large $(y_i-y_j)\to \pm \infty$. 

The Centauro algorithm defines the following distance measure 
%%%
\begin{align} \label{eq:dijcen}
    d_{ij}&=\left[
    (\Delta f_{ij})^2+2f_i f_j (1-\cos\Delta\phi_{ij})
    \right]/R^2,\quad d_{iB}=1
    \,,
\end{align}
%%%    
where $\Delta f_{ij}$ and $ \Delta\phi_{ij}$ are differences between $f_i$ and $f_j$ and $\phi_i$ and $\phi_j$, respectively. The function $f_i$ is a function of an angular variable $\bar\eta_i$ of particle $i$
%%%    
\begin{align} \label{eq:fi}
    f_i &= f(\bar{\eta}_i)\,,\qquad\qquad
    \bar{\eta}_i
    = 2\frac{|\mathbf{p}_{iT}|}{\bnz\mcdot p_i}
    \,,
\end{align}
%%%
 where $\bar{\eta}$ diverges in forward region as $\bnz\mcdot p_i\to 0$ and thus prevents jets from enclosing the proton beam direction, while it decreases as particles become closer in backward region $ {\bnz\mcdot p_i}\gg |\mathbf{p}_{iT}|$. 
For the function $f$, we take the simplest form in \cite{Arratia:2020ssx}
%%%
 \be\label{eq:feta}
 f(\bar\eta)=\bar\eta
\,. 
\ee
%%%
Then, $f_1$ and $f_2$ and the distance $d_{12}$ are given by
%%%
 \begin{align}
  \label{eq:f1f2}
 f_1 &=2\sqrt{\frac{1-z}{z}}\sqrt{\frac{v}{1-v}}\,,
 \qquad\qquad
 f_2 =2\sqrt{\frac{1-z}{z}}\sqrt{\frac{1-v}{v}}
  \,,\\
 \label{eq:d12cen}
  d_{12} &=\frac{(f_1+f_2)^2}{R^2}
  =    \frac{4}{R^2}\frac{1-z}{z}\frac{1}{v(1-v)}
 \,,
 \end{align}
 %%%
where $R$ is a radius parameter of order $\cO(1)$. If $d_\text{min}$ is a $d_{iB}$, we have the same jet as the anti-$k_T$ jet. If $d_\text{min}$ is a $d_{12}$, we have a jet with momentum ${\mathbf p}_1+{\mathbf p}_2$, which is not allowed in the anti-$k_T$ algorithm. The condition for two particles merging into a single jet is $d_{12}<d_{iB}$, which indicates the following region
 %%%
 \begin{align} \label{eq:ThIV}
  \Thfour(z,v)=\theta\left(z-z_c(R)\right)\theta\left(v-v_{-}(z,R)\right)\theta\left(-v+v_{+}(z,R)\right)\,,
 \end{align}
 %%% 
where $v_\pm$ are upper and lower bounds of the variable $v$
and $v_\pm=1/2$ at the value $z=z_c(R)$
%%%
\begin{align} \label{eq:vpm-zc}
    v_\pm(z,R)=\frac{1}{2}\pm\frac{1}{2}\sqrt{1-\frac{1-z}{z}\frac{16}{R^2}}\,,
   \quad  
 z_c(R)=\frac{16}{16+R^2}\,.
\end{align}
%%%

In this region, the jet momentum is $\mathbf{p}_1+\mathbf{p}_2$ and  $q_J$ always is in the opposite to the proton direction
%%%
\begin{align}\label{eq:qjct}
    q_J^\text{ct}
=|\mathbf{p}_1+\mathbf{p}_2|(1,\hat{n}_{12})
=Q\frac{2z-1}{z}\,\frac{\nz}{2}
\,,
\end{align}
%%%
where $\hat{n}_{12}=(\mathbf{p}_1+\mathbf{p}_2)/|\mathbf{p}_1+\mathbf{p}_2|$.
Otherwise $q_J^\text{ct}=q_J^\text{kt}$ is one of $p_1$ and $p_2$ as in \eq{qjkt}.
The right panel in \fig{vz-space} indicates the phase space for 1-jettiness with the Centauro algorithm. 
The bounds of region IV for $R=1,2$ are shown in the figure in red curves. The region reduces to zero as $R\to 0$ and invades the regions I and II for $R>2$. In many experiments, $R$ is taken to be less than or, equal to 2 and the phase space in this range is also relatively simple. So, we constrain the region of our interest to be
%%%
\be\label{eq:Rle2}
R\le 2
\,.\ee
%%%

Then, in the region IV the value of $\tauct$ can be computed with Centauro-jet axis in \eq{qjct} and in the regions I, II, and III $\tauct=\taujt$. Now $\tauct$ in any region can be expressed as
%%%
\begin{align}\label{eq:tauct}
    \tauct=\Thfour\frac{(2z-1)(1-z)}{z^2}+(1-\Thfour)\taujt 
\,.
\end{align}
%%%
Note that $\tauct$ is always positive in the region IV as well as the other regions because the minimum value of $z$ is $z_c(R)\geq 4/5$ in the range given by \eq{Rle2}.

%%%%%%%%%%%%%%%%%%%%%%%%%%%%%%%%%%%%%%%%%%%%%%%%%%%%%%
\section{Analytic cross section at order $\as$} \label{sec:xsec}
%%%%%%%%%%%%%%%%%%%%%%%%%%%%%%%%%%%%%%%%%%%%%%%%%%%%%%

In this section we summarize the analytic result for 1-jettiness cross section in DIS computed at the first order in $\as$. Details of the calculations are given in \appx{detail}. Here, our results are given in the form of cumulative cross section obtained by integrating the differential cross section
%%%
\be\label{eq:sigma}
\sigmac(\tau,x,Q) = \int_0^\tau d\tau' \frac{d\sigma}{dx\, dQ^2\, d\tau'}
\,.
\ee
%%%
The DIS cross section is conventionally decomposed into the structure functions $F_i$
%%%
\begin{align}
\label{eq:sigma-Fi}
\sigmac(\tau,x,Q)
 &=
\frac{4\pi \alpha^2}{Q^4}\left[ (1+(1-y)^2)F_1 +\frac{1-y}{x}F_L\right]
\,.
\end{align}
%%%
This can be expressed in terms of $F_1,F_2$ by using the relation between structure functions $F_2=F_L+2x F_1$. 
Note that the structure functions are the functions of $\tau$ as well as $x$ and $Q$.
At the point $\tau=\tau_\text{max}$, the cumulative functions $\sigmac$ and $F_i$ reduce to the inclusive cross section and inclusive structure functions, respectively. 
The $F_i$ are coefficients of basis tensors that decompose a current-current correlator called the hadronic tensor $W^{\mu\nu}$ defined in \eq{Wtensor} and one can read off the individual coefficients by projecting  with orthogonal tensors like $P_\mu P_\nu W^{\mu\nu}$ and $g_{\mu\nu}W^{\mu\nu}$ as shown in \eq{Fi}.

The structure functions are written in terms of these projected correlators
%%%
\begin{align}\label{eq:F1FL}
F_1 &=\sum_{i=q,\bar q,g}  (A_i+B_i) 
\,,\\
F_L &=\sum_{i=q,\bar q,g} 4x \, A_i
\nn\,,
\end{align}
%%%
where $A_i$ and $B_i$ are respectively equivalent to $P_\mu P_\nu W^{\mu\nu}$ and $g_{\mu\nu}W^{\mu\nu}$ up to multiplicative factors that can be found in \eqs{ppW-proton}{gW-proton}.
They contain logarithmic terms (singular) that were obtained
by the factorization formula in \cite{Kang:2013nha,Kang:2012zr}.
Non-logarithmic terms (nonsingular) are obtained in this study by fixed-order QCD calculations. 
Each of $A,B$ can be written like $A=A^\sing+A^\ns$ and $B=B^\sing+B^\ns$.
For the completeness we copy and paste the singular parts
%%%
\begin{align} 
A_{q}^\sing &= A^\sing_g=0
\,,
\nn\\
B_q^\sing &=
\sum_f Q_f^2 \biggl\{  
f_q(x) \bigg[\frac12
 -   \frac{\as C_F}{4\pi}  \biggl(\frac92+\frac{\pi^2}{3} + 3 \ln\tau +2\ln^2\tau\biggr) \bigg]
\nn\\&\quad
+ \frac{\as C_F}{4\pi} \int _x^1 \frac{dz}{z} f_q\left(\frac{x}{z}\right) 
\biggl[\cL_1(1-z) \,(1+z^2)+1-z + P_{qq}(z)\ln \frac{Q^2 \tau}{\mu^2z}\biggr]  
\biggr\}
\,,
\nn\\
B_g^\sing&=
 \sum_f Q_f^2 \frac{\as T_F}{2\pi} 
\int_x^1 \frac{dz}{z}f_g\left(\frac{x}{z}\right) \Biggl[ 1-P_{qg}(z) 
+ P_{qg}(z)\ln  \left( \frac{Q^2\tau}{\mu^2}\frac{1-z}{z} \right) \Biggr] 
\,,
\label{eq:AsBs}
\end{align}
%%%
where $P_{qq}(z)$ and $P_{qg}(z)$ are the splitting functions and $\cL_n(1-z)$ is a plus distribution 
%%%
\begin{align}\label{eq:Pqqqg}
P_{qq}(z)&=\bigg[\theta(1-z)\frac{1+z^2}{1-z}\bigg]_+
\,,
\nn\\
P_{qg}(z)&= [(1-z)^2+z^2]\,,
\nn\\
\cL_n(1-z)&=\bigg[\frac{\theta(1-z)\ln^n (1-z)}{1-z}\bigg]_+
\,.
\end{align}
%%%
Note that the anti-quark contributions $A_{\bar q}$ and $B_{\bar q}$ are the same as $A_q$ and $B_q$ except for the quark PDF replaced by the anti-quark PDF.

The nonsingular parts are computed in \appx{detail}. The final expressions are given by
%%%
\begin{align}
\Ans_q &=
\sum_f  Q_f^2 \frac{\as C_F}{4\pi}\biggl\{(2\tau-1)\, \Theta_0 \int_x^{\frac{1}{1+\tau}} dz \, f_{q} \left(\frac{x}{z}\right)  + \int_x^1 dz\, f_{q}\left(\frac{x}{z}\right)\biggr\}
\,,\nn\\
\Ans_g &=
\sum_f Q_f^2 \frac{\as T_F}{\pi} \biggl\{(2\tau-1)\, \Theta_0 \int_x^{\frac{1}{1+\tau}} dz f_g\left(\frac{x}{z}\right)  (1-z)  + \int_x^1 dz\, f_g\left(\frac{x}{z}\right) (1-z)\biggr\}
\,,\nn\\
\Bns_q &=
\sum_f Q_f^2 \frac{\as C_F}{4\pi}
 \Biggl( \Theta_0
\biggl\{
\int_{x}^{\frac{1}{1+\tau}} \frac{dz}{z} f_q\left(\frac{x}{z}\right) \Bigl[ \frac{1-4z}{1-z}(\tau - 1/2)- P_{qq}(z)\ln\frac{1-\tau}{\tau}\Bigr]\biggr\}
\nn \\ & \quad
+\int^1_x \frac{dz}{z}\,f_q\left(\frac{x}{z}\right)\bigg[\cL_0(1-z)\frac{1-4z}{2}-P_{qq}(z)\ln\tau \bigg]+f_q(x)\big(2\ln^2\tau+3\ln\tau \big)
\bigg)
\,,\nn\\
\Bns_g &=
 \sum_f Q_f^2 \frac{\as T_F}{2\pi} \biggl\{
 \Theta_0 \int^{\onet}_{x}\frac{dz}{z}\,f_g\left(\frac{x}{z}\right)
 \bigg[1-2\tau - P_{qg}(z)\,\ln\frac{1-\tau}{\tau}\bigg]
 \nn\\
 &\qquad\qquad\qquad\qquad\qquad\qquad\qquad\qquad\qquad
 -\int^1_x \frac{dz}{z}\, f_g\left(\frac{x}{z}\right)\big[1+P_{qg}(z)\ln\tau  \big]
  \biggr\}
\,,
\label{eq:AnsBns}
\end{align}
%%% 
where $\Theta_0$ represents physical region of $\tau$ for a given value of $x$ 
%%%
\begin{align}\label{eq:theta0}
\Theta_0(\tau,x)&\equiv\theta(\tau)\,\theta\bigg(-\tau+\frac12\bigg)\,\theta\bigg(-\tau+\frac{1-x}{x}\bigg)
\,.
\end{align}
%%%
In comparison to $\taub$, the singular and nonsingular parts of $\taujt$ have many terms in common with those of $\taub$ in \cite{Kang:2014qba}. Their differences are summarized in \appx{diff}.

As shown in \eq{tauct}, $\tauct$ is different
from $\taujt$ in the region IV. We take their differences in the region and denote them by $\delta A_i$ and $\delta B_i$. 
Then, the structure functions for $\tauct$ are obtained by replacing $A$ by $A+\delta A$ and $B$ by $B+\delta B$ in \eq{F1FL}. 
Their final expressions are given by
%%%
\begin{align} 
   \dAns_q &=- \sum_f Q^2_f\frac{\as C_F}{4\pi}\,
    \theta_\text{ct}(\tau)\intd  dz\, f_q \left(\frac{x}{z}\right)
    r(z,R)
    \,,\nn\\
   \dAns_g &=- \sum_f Q^2_f \frac{\as T_F}{\pi}\, 
   \theta_\text{ct}(\tau)
    \intd  dz\, f_g \left(\frac{x}{z}\right)
    (1-z)r(z,R)
    \,,\nn\\
    \dBns_q &=- \sum_f Q^2_f\frac{\as C_F}{4\pi}\,
   \theta_\text{ct}(\tau)
   \intd  \frac{dz}{z}\, f_q \left(\frac{x}{z}\right)
    \left\{\frac{1-4z}{2(1-z)}r(z,R)+\frac{1+z^2}{1-z}\ln\frac{1+r(z,R)}{1-r(z,R)}   \right\}
    \,,\nn\\
   \dBns_g &= -\sum_f Q^2_f\frac{\as T_F}{2\pi}
   \theta_\text{ct}(\tau)
   \intd \frac{dz}{z}\,f_g \left(\frac{x}{z}\right)
   \left\{ -r(z,R)+P_{qg}(z)\ln\frac{1+r(z,R)}{1-r(z,R)}   \right\}
    \,,
   \label{eq:dABns}
\end{align}
where $\theta_\text{ct}$ is upper limit of $\tau$ 
%%%
\be\label{eq:th_ct}
\theta_\text{ct}(\tau)=
\theta\left(-\tau+\frac{R^2}{16}\right)
\,,
\ee
%%%
and the parameters $z_{\text{jt},\text{ct}}$ and $r(z,R)$ are given by
%%%
\begin{align}
\label{eq:zjt}
\zjt&=\max \left\{x,\frac{1}{1+\tau}\right\}
\,,\\ \label{eq:zct}
\zct&=\max \left\{x\,, \frac{16}{16+R^2}\,, \frac{3+\sqrt{1-4\tau}}{2(2+\tau)} \right\} \,,
\\
\label{eq:r}
r(z,R)&=v_{+}-v_{-}
 =\sqrt{1-\frac{1-z}{z}\frac{16}{R^2}}
\,.
\end{align}
%%%
The differential distributions can be obtained by differentiating \eqs{AnsBns}{dABns}.
We also give their explicit expressions for $\taujt$ in \eqs{cAq}{cAg} and \eqs{cBq}{cBg}
and for $\tauct$ in \eq{delcAcB}.

In addition,  by expanding $A^{\text{ns}}_{q}$, $A^{\text{ns}}_g$, $B^{\text{ns}}_g$ and $B^{\text{ns}}_{q}$  in the $\tau \rightarrow 0$ limit we can obtain the NLP term which is the power correction to the singular terms in \eq{AsBs}. At the leading power the corrections contain terms like $\tau$ and $\tau\ln \tau$, which is suppressed by $\tau$ compared to the singular terms.
The NLP obtained from \eq{AnsBns} can be expressed as
%%%
\begin{align}
A^{\text{ns}}_{q}
 \big|_{\tau\rightarrow 0}
& =  \sum_f  Q_f^2 \frac{\as C_F}{4\pi}\,\tau
\biggl\{
f_{q}(x)+2\int^1_x dz\, f_{q}\left(\frac{x}{z}\right)
\biggr\} +\cO(\tau^2)
\,,\nn\\
A^{\text{ns}}_{g}\big|_{\tau\rightarrow 0}
& = \sum_f Q_f^2 \frac{\as T_F}{\pi}\,\tau
\biggl\{
2\int^1_x dz\, f_{g}\left(\frac{x}{z}\right)(1-z)
\biggr\} +\cO(\tau^2)\,,\nn\\
B^{\text{ns}}_{q}\big|_{\tau\rightarrow 0}
&=
\sum_{f} Q_f^2 \frac{\as C_F}{4\pi}\,
\tau\biggl\{
\bigg(\frac12+3\ln{\tau}\bigg)\, f_q(x)-\bigg(\frac{3}{2}+2\ln{\tau}\bigg)\,x \,f^{'}_{q}(x)\nn\\
&\qquad\qquad\qquad+\,\int^1_x\frac{dz}{z}\,f_q\,\left(\frac{x}{z}\right)
\bigg[
(1-4z)\cL_0(1-z) + P_{qq}(z)
\bigg]
\biggr\} + \cO(\tau^2)
\nn\,,\\
B^{\text{ns}}_g\big|_{\tau\rightarrow 0}
&= 
\sum_f Q_f^2 \frac{\as T_F}{2\pi} \,\tau
\biggl\{
-
f_g\left(x\right)
\big(
1+\ln\tau
\big)
+\int^1_x\frac{dz}{z} f_g\,\left(\frac{x}{z}\right)
\big[
P_{qg}(z)-2
\big] 
\biggr\} +\cO(\tau^2)
\,,
\label{eq:pc-ABns}
\end{align}
%%%
where $f'(x)= df(x) /dx$.

In expansions of \eq{dABns}, we assume that $\tau < R^2/16$ then, $z_\text{ct}=1-\tau-\tau^3+\cO(\tau^4)$. Otherwise, \eq{dABns} vanishes. At $\cO(\tau)$ the power corrections are zero except for $\delta B^{\text{ns}}_q$. So, expanding up to first nonzero correction we have
%%%
\begin{align}
    \delta A^{\text{ns}}_q\big|_{\tau\rightarrow 0} &= -\sum_f Q^2_f\frac{\as C_F}{4\pi}\,
    \tau^2\,f_q(x) + \cO(\tau^3)
    \,,\nn\\
    \delta A^{\text{ns}}_g\big|_{\tau\rightarrow 0} &= -\sum_f Q^2_f\frac{\as T_F}{\pi}\, 
    \tau^3\,f_g(x) + \cO(\tau^4)
   \,,\nn\\
    \delta B^{\text{ns}}_q\big|_{\tau\rightarrow 0} &=
    -\sum_f Q^2_f\frac{\as C_F}{4\pi}
    \tau\bigg(-\frac32+2\ln\frac{R^2}{4\tau} \bigg)\,f_q(x) + \cO(\tau^2)
%    \tau[-\frac32+2\ln\frac{R^2}{4\tau}-\tau(\ln{\frac{R^2}{4\tau}}+\frac{4}{R}-\frac{1}{4})]\,f_q(x) + \cO(\tau^3)
    \,,\nn\\
        \delta B^{\text{ns}}_g\big|_{\tau\rightarrow 0} &= -\sum_f Q^2_f\frac{\as T_F}{2\pi}
    \tau^2 \bigg(\ln\frac{R^2}{4\tau}-1\bigg)\, f_g(x)
    + \cO(\tau^3)
\,.
\label{eq:pc-dABns}
\end{align}
%%%

It is worth to note that PDFs derivatives appear in the NLP results which should be identified in a factorization at the subleading power and  $\cO(\sqrt{\tau_1})$ corrections are absent for any of the jet algorithms presented since power corrections of 1-jettiness scales like $\cO(\tau_1^n)$ with integer $n$. 
 Another feature to note is that the appearance of a jet radius dependence in \eq{pc-dABns} is expected from the effective theory analysis \cite{Ellis:2010rw} that implies that power corrections are sensitive to the jet algorithms and the logarithmic structure due to the phase-space cutoff typically appear at the leading power in jet substructure observables \cite{Hornig:2016ahz} and in transverse vetos \cite{Hornig:2017pud}.

In the context of high-order calculations in perturbative QCD,
the observable N-jettiness is used to control and to subtract the infrared singularities in numerical computation \cite{Boughezal:2015dva,Boughezal:2015aha,Gaunt:2015pea}, the results in \eqs{pc-ABns}{pc-dABns} are essential ingredients required to improve the subtraction accuracy to subleading order \cite{Moult:2016fqy,Ebert:2018lzn}, while the singular part in \eq{AsBs} is for subtraction at the leading order. 
The results also serve as an important crosscheck to be reproduced by effective field theory approach using the factorization at subleading power on the way to the higher-order in $\as$. 
It is worth to point out an interesting observation in our results. The algorithms, 1-jettiness and anti-$k_t$, are the same at the order $\as$ hence, their subtractions to arbitrary accuracy are identical at this order. On the hand, the jet algorithms, anti-$k_t$ and Centauro, are different in the subtraction at the subleading power because their difference in \eq{pc-dABns} contains non-vanishing $\tau \ln \tau$ terms.

%%%%%%%%%%%%%%%%%%%%%%%%%%%%%%%%%%%%%%%%%%%%%%%%%%%%%%
\section{Numerical results} \label{sec:num}
%%%%%%%%%%%%%%%%%%%%%%%%%%%%%%%%%%%%%%%%%%%%%%%%%%%%%%
In this section, we show our numerical results obtained by using analytic 1-jettiness cross sections at order $\as$ in \sec{xsec}. We compare singular and nonsingular parts at different values of $x$ and $Q$. Numerical difference between $\taujt$ and $\tauct$ are also presented for several values of $R$. We also compare our result to $\taub$ result given in \cite{Kang:2014qba}. 
The crossing point between singular and nonsingular parts implies a boundary between fixed-order and resummation regions and the point as a function of $x$ at several values of $Q$ is presented.
In choosing the values for the kinematic parameters $x$ and $Q$ in the section, we considered the region that can be studied in future experiments such as EIC or $\EICC$.
%%%%%%%%%%%%%%%%%%%%%%
\begin{figure}[t!]
\vspace{-1ex}
\includegraphics[width=.47\columnwidth]{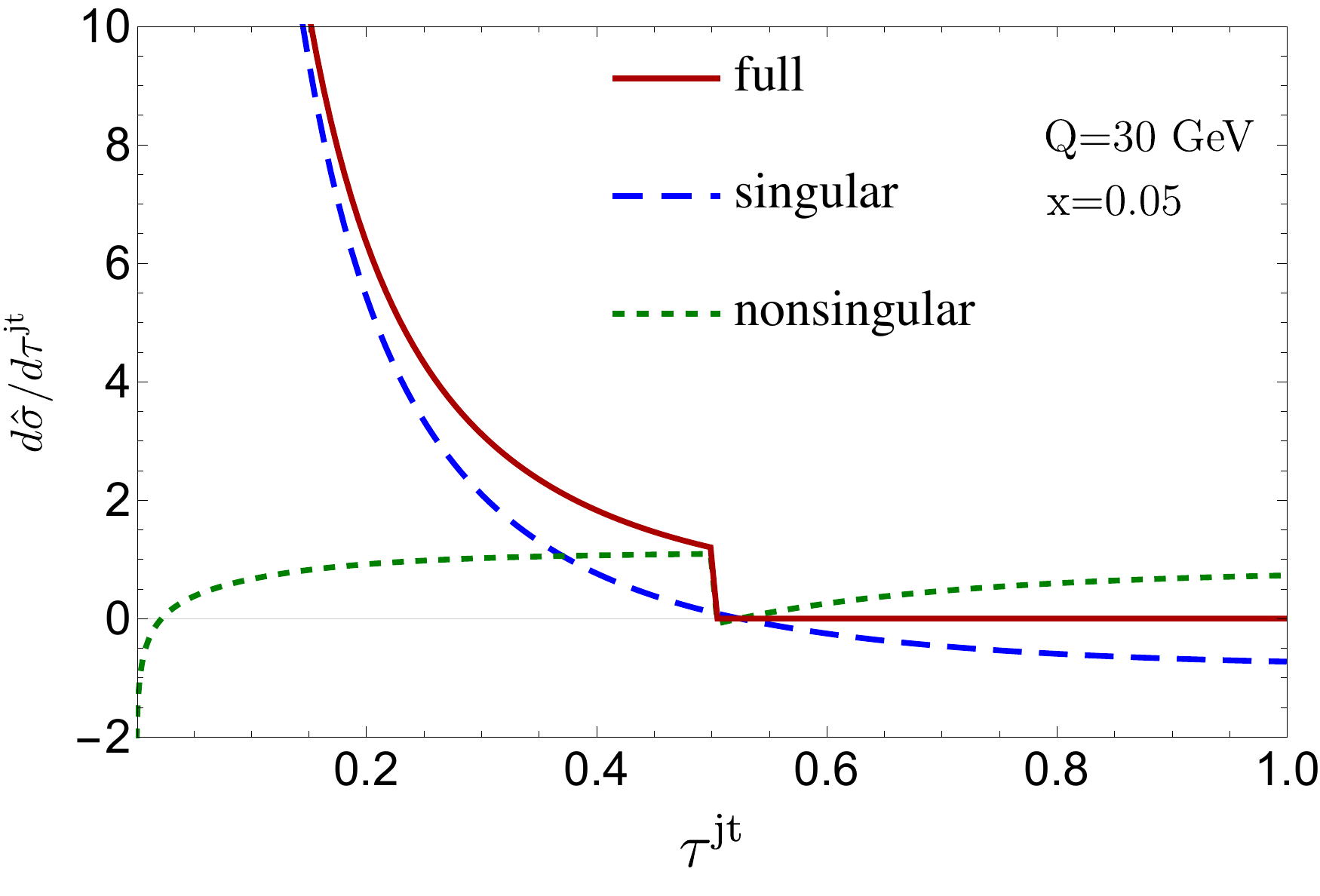}\hspace{4ex}
\includegraphics[width=.49\columnwidth]{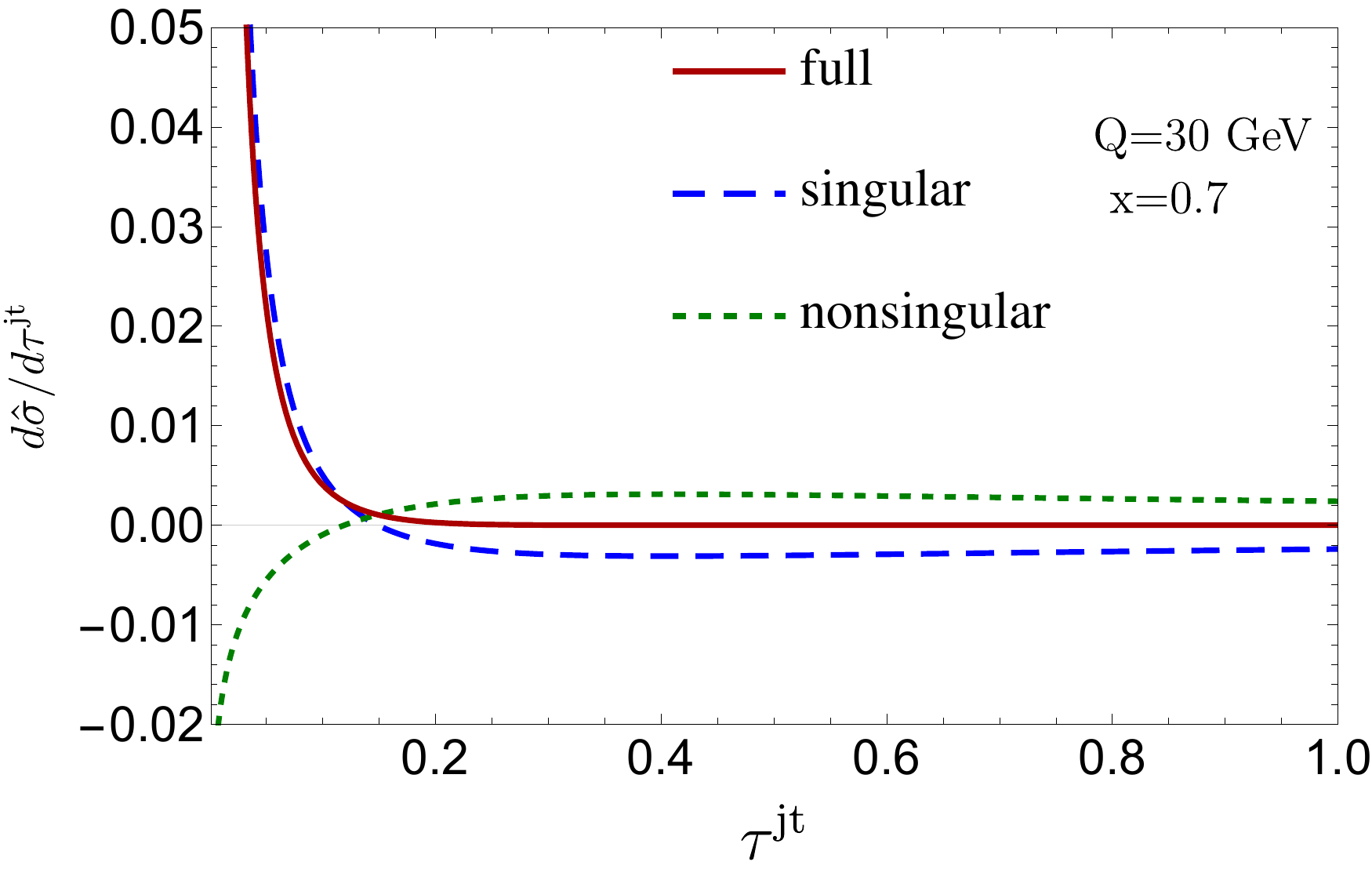} 
\vspace{-1em}
	\caption{ Differential $\taujt$ cross section at $Q=30$ GeV and $x=0.05$ and 0.7.
	Singular (blue dashed), nonsingular (green dotted), and their full (red solid) distributions 
  \label{fig:singns}
	}
\end{figure}
%%%%%%%%%%%%%%%%%%%%%%

Main achievement of this paper is the analytic expressions of nonsingular part of $\taujt$ and $\tauct$. 
In \fig{singns} we show the differential $\taujt$ cross sections where singular and nonsingular parts as well as full fixed-order results are presented at two values of $x$ with a fixed value of $Q$. The result at different values of $Q$ shows the similar pattern.
The cross section shown in the plots is normalized as
%%%
\be
\frac{d\hat\sigma}{d\tau}= \frac{1}{\sigma_0} \frac{d\sigma}{d\tau}
\,,
\ee
%%%
where $\sigma_0=2\pi\alpha^2\left[ 1+(1-y)^2\right]/Q^4$.
Note that $\taujt$ for 2-body final state vanishes at $\tau=1/2$ when $x<2/3$ and at $\tau=(1-x)/x$ when $x>2/3$. Beyond the point the singular and nonsingular cancel in the sum and the full cross section is zero.

%%%%%%%%%%%%%%%%%%%%%%
\begin{figure}[thb]
\begin{center}
\includegraphics[width=.6\columnwidth]{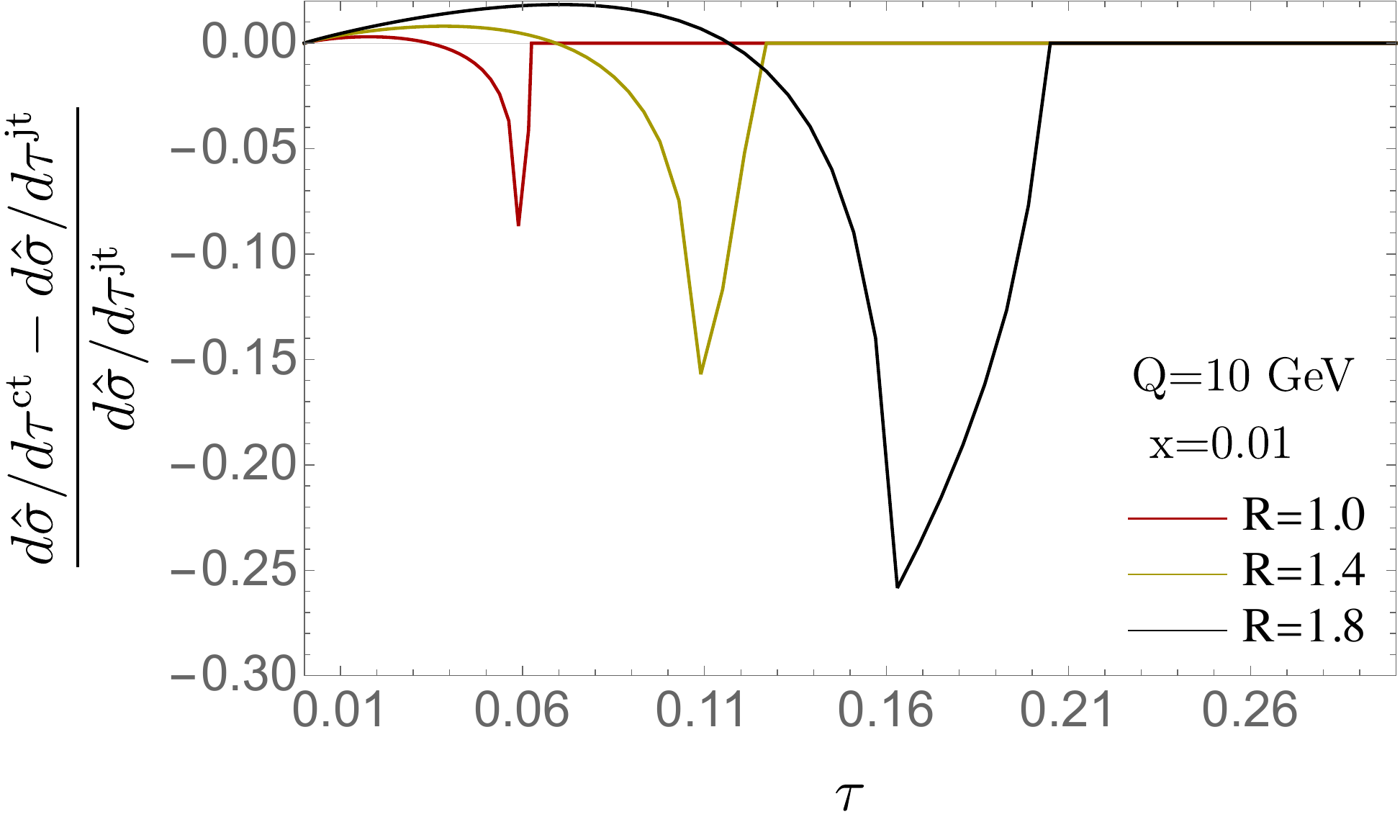}
\end{center}
\vspace{-4ex}
\caption{Relative difference between $\tauct$ and $\taujt$ distributions with three values of $R=1.0\,, 1.4\,, 1.8$ at $x=0.01$ and $Q=10$ GeV.
\label{fig:ct-th}}
\end{figure}
%%%%%%%%%%%%%%%%%%%%%%

About the other variant $\tauct$, its singular part is the same as that of $\taujt$ and the nonsingular is similar to  $\taujt$. Instead of showing the similar style of plots for $\tauct$, we compare difference between $\taujt$ and $\tauct$ in \fig{ct-th} by using the expression in \eq{dABns}. The size of difference increases with $R$ increasing and it becomes as large as 25\% when $R=1.8$. With smaller values of $R$ the difference quickly reduces to zero and this is consistent with the $R$ dependence of the region IV in \fig{vz-space} left panel, where the region shrinks to zero as $R\to 0$. A feature of sharp turning upwards in the plot is associated with the contributions from $\tauct$ vanishing in $\tau\ge (16-R^2)R^2/256$. Then, when the other contribution from $\taujt$ vanishes in $\tau\ge R^2/16$ the difference becomes zero as shown in the figure.

%%%%%%%%%%%%%%%%%%%%%%
\begin{figure}[thb]
	\begin{center}
		\includegraphics[scale=.32]{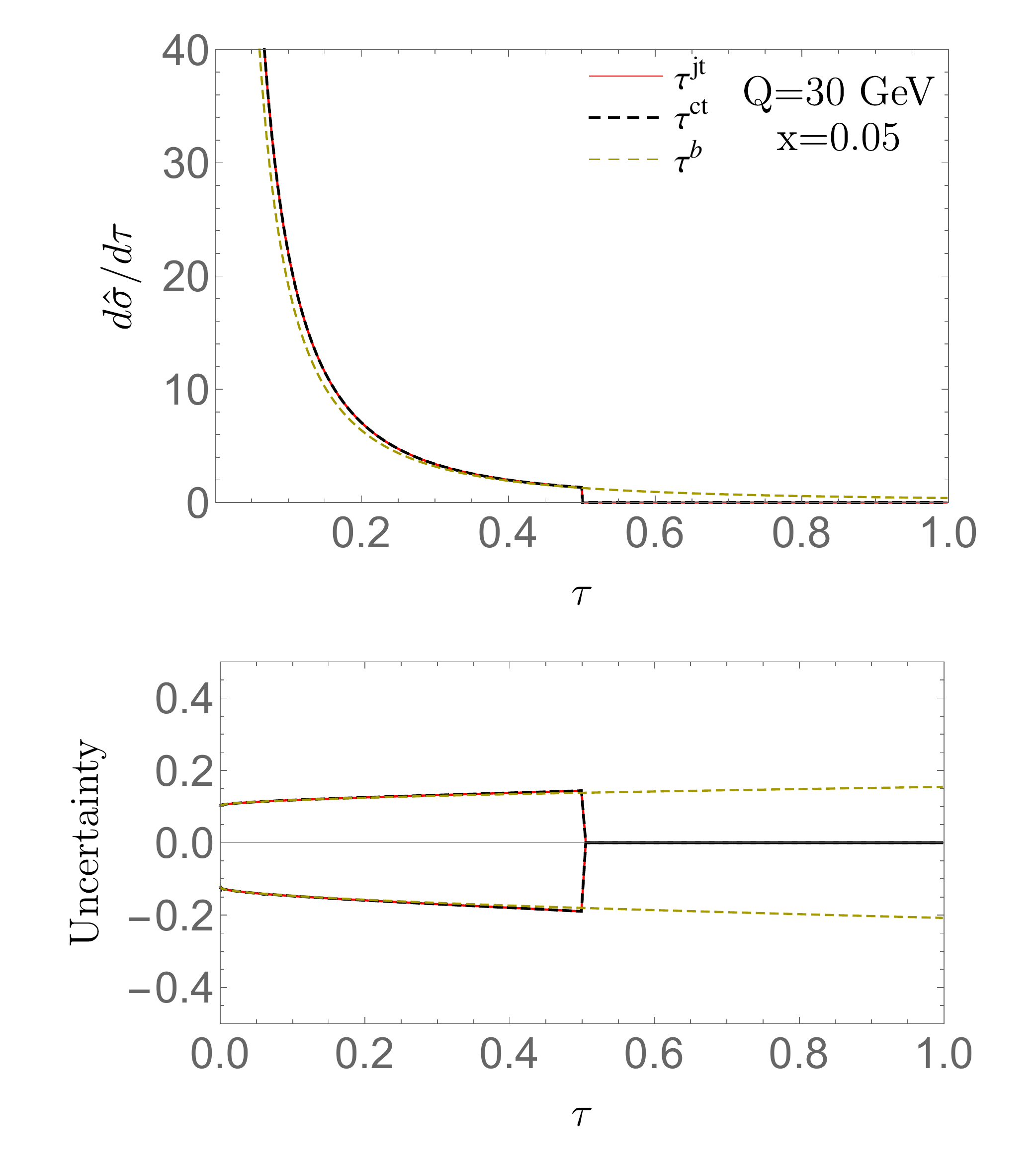}
		\includegraphics[scale=.32]{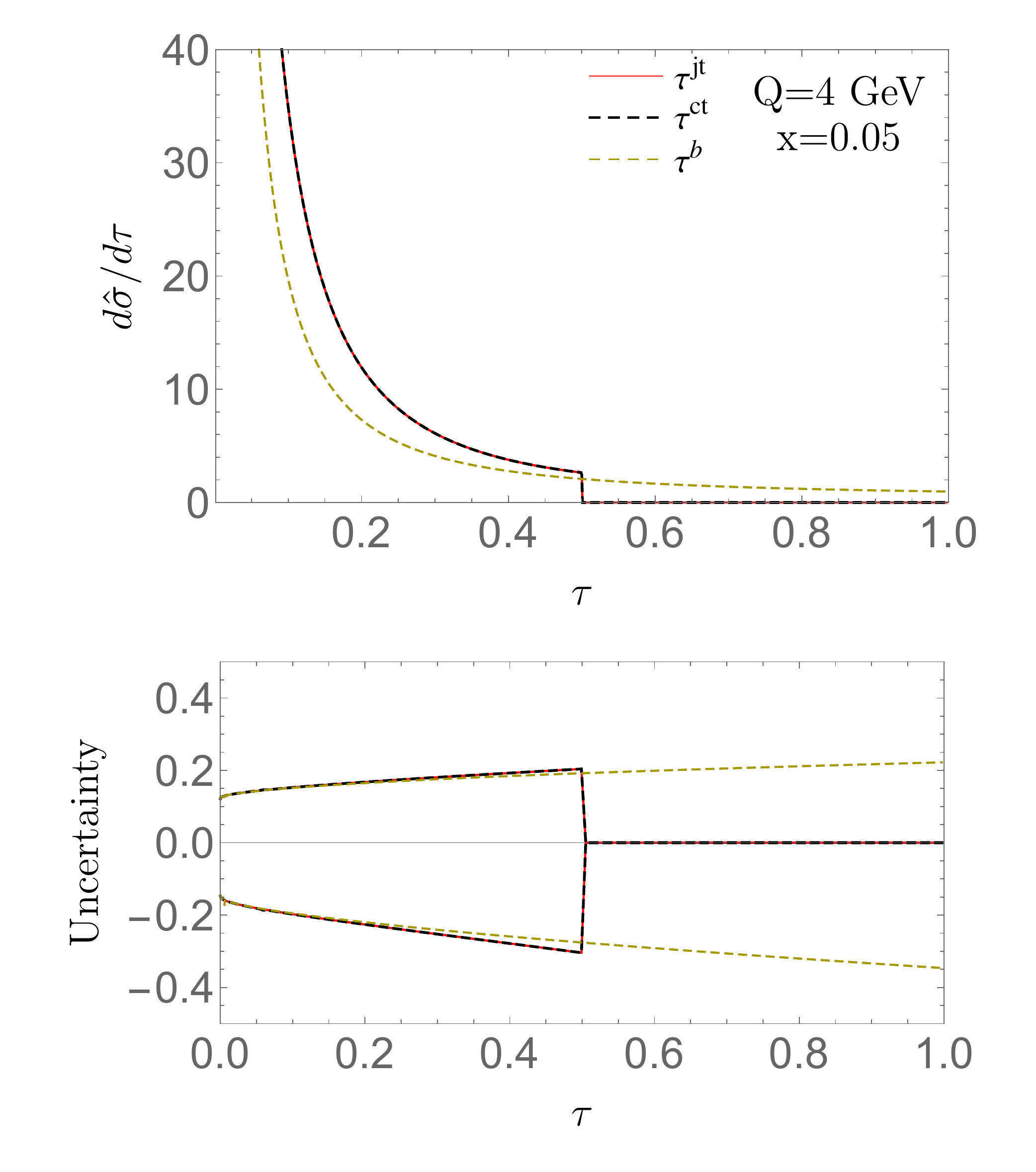}  
	\end{center}\vspace{-4ex}
	\caption{Upper two figures are $\taujt$ and $\tauct$ distribution in comparison to $\taub$ at $Q=30$ GeV (left) and 4 GeV (right) with fixed values of $x=0.05$ and $R=1$ and lower ones are corresponding uncertainties.
	\label{fig:taub}
	}
\end{figure}
%%%%%%%%%%%%%%%%%%%%%%
In \fig{taub} we compare our result to another version of 1-jettiness called $\taub$ in \cite{Kang:2014qba}.
Because of difference in factorized formula between them, their singular parts are different \cite{Kang:2013nha}.
However the difference in singular part is proportional to $\delta(\tau)$ and is not visible in a differential distribution like \fig{taub}. Therefore, the difference from $\taub$ shown in the plot is from nonsingular parts and the size of difference is larger at lower value of $Q$. Note that  as shown in the plot $\taub$ does not vanish beyond $\tau=1/2$ because the maximum of $\taub$ is 1, while it is 1/2 for $\taujt$ and $\tauct$. For the uncertainty figure, the uncertainties are obtained by varying the renormalization scale from $\mu=Q$ by a factor of two up and down $(d\hat\sigma(Q)-d\hat\sigma(2Q))/d\hat\sigma(Q)$ and $(d\hat\sigma(Q)-d\hat\sigma(Q/2))/d\hat\sigma(Q)$ as upper and lower boundary, respectively. We can find that $\taujt$ and $\tauct$ share almost same uncertainty. They only have slightly difference with each other within  $\tau=1/2$. This uncertainty increases with the decrease of the values of $Q$. We also compare uncertainties for $\tauct$ with different $R$ value and find that the effect is around $0.01\%$ changes.

\begin{figure}[htb]
	\begin{center}
		\includegraphics[scale=.40]{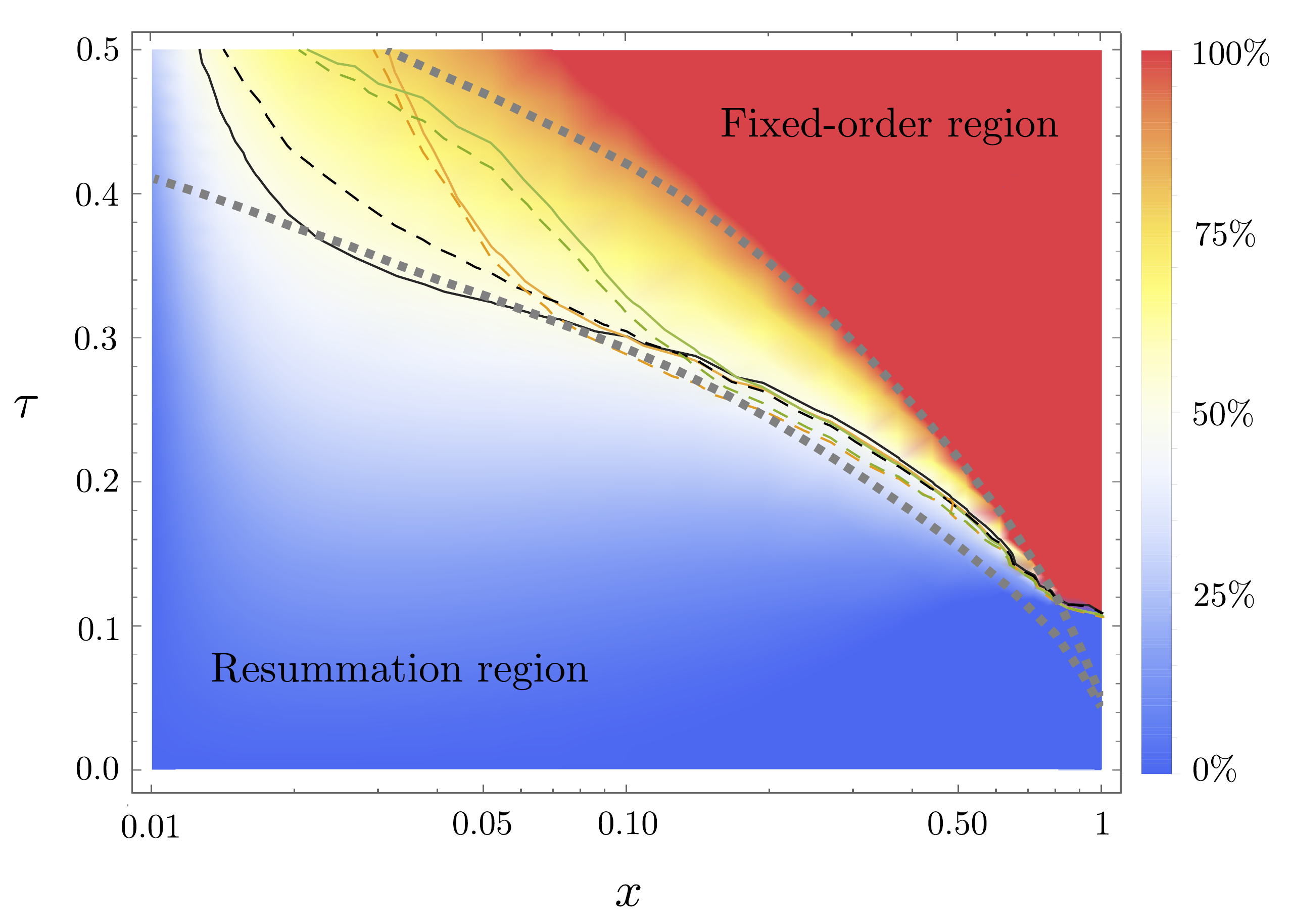} 
	\end{center}\vspace{-4ex}
\caption{Values of $\taujt$ (solid lines) and $\taub$ (dashed lines) when the singular and nonsingular parts cross each other as a function of $x$ at $Q=15$ GeV (black), $30$ GeV (orange), and $45$ GeV (green).
The color density plot represents the portion of the contributions from the nonsingular part to the total differential 
cross section at $Q=15$ GeV for $\taujt$.
An approximate crossing (dotted gray lines) are obtained by making use of 
\eq{taucross} with the values of $\cns$ between  $0$ and $-\beta(x)$.
	\label{fig:crossing}
	}
\end{figure}

The crossing point where singular and nonsingular parts meet each other can be understood as  a boundary between fixed-order and resummation regions. In the fixed-order region,  the singular part is smaller than the nonsingular part hence, an ordinary fixed-order QCD result is valid, while in the resummation region, the singular part is larger than nonsingular part  due to increasing logarithms and resummation of those logarithms is necessary. In \fig{crossing} the crossing points in $\tau$ as a function of $x$ are shown and the results for $\taujt$ (solid) and $\taub$ (dashed) are similar as implied from \fig{taub}. 

The color density plot in \fig{crossing} represents a relative size of the nonsingluar part to the full differential cross section at $Q=15$ GeV for $\taujt$. An absolute values are taken for simplicity.
In the blue left and lower corner the nonsingular is small while the singular dominates the cross section. 
In the light-colored region the singular and nonsingular are comparable to each other. 
Finally, in deep-red right and upper corner unphysical singular and nonsingular are largely cancelled to give the full cross section. In the region their absolute magnitudes easily become greater than full cross section and the deep-red region should be understood as 100\% or, greater relative to the full cross section.

An important feature in the plot is that the resummation region increases with decreasing value of $x$ and the region gets close to the maximum of $\taujt$ near $x=0.01$ while it does near $x=10^{-5}$ for $\taub$. These crossing points imply when the resummation should be turnoff in $\tau$ spectrum and specifically in the scale profile function \cite{Kang:2013nha,Kang:2014qba} the points can be taken to be the value of a parameter $t_2$ which is the point where the resummation begins to be turned off.

%%%%%%%%%%%%%%%%%%%%%%
%%%%%%%%%%%%%%%%%%%%%%

In order to understand the behavior shown in \fig{crossing}, we take a following form
%%%
\be\label{eq:albecns}
 \alpha(x) \frac{\ln{\tau}}{\tau}+\beta(x) \frac{1}{\tau}= \cns
\,,
\ee
%%%
where the left side is from the singular part and the coefficients $\alpha(x)$ and $\beta(x)$ are obtained by differentiating \eq{AsBs}.
%%%
\begin{align}\label{eq:albe}
 \alpha(x) &= -8 C_F \sum_{f} Q_f^2 f_q(x)\,,
 \nn\\
 \beta(x) &= \sum_f Q_f^2 \left[ -6C_F f_q(x)+2C_F \int_x^1 \frac{dz}{z}P_{qq}(z) f_q\left(\frac{x}{z}\right)+2T_F \int_x^1 \frac{dz}{z} P_{qg}(z) f_g\left(\frac{x}{z}\right)\right]
\,.
\end{align}
%%%
On the right-hand side of \eq{albecns}, the parameter $\cns$ corresponds to the nonsingular part in \eq{AnsBns} multiplied by a proper normalization factor.
Instead of using the known result we assume $\cns$ is a unknown constant of $\tau$ and would like to find an approximate solution to \eq{albecns} by using an iterative approach.
First, we set $\cns$ to zero, then find the solution to \eq{albecns} is $\tau=\exp[{-\beta(x)/\alpha(x)}]$. Second by introducing a small correction term $\delta (x)$ to the solution for zero $\cns$ as in \eq{taucross}, we solve \eq{albecns} for $\delta(x)$ with nonzero $\cns$. The approximate solution can be expressed as
%%%
\be\label{eq:taucross}
\tau_\text{cross}(x,\cns)=e^{-\frac{\beta(x)}{\alpha(x)}+\delta(x,\cns)}
\,,
\ee
%%%
where $\delta(x)=(\cns/\alpha(x)) \exp[{-\beta(x)/\alpha(x)}]$. We can find the values of $\cns$ that makes the approximate solution fitted to the curves in \fig{crossing} obtained using the known nonsingular part. We find that $\cns$ between 
$0$ and $-1$ times $\beta(x)$, represented by the grey dotted lines in \fig{crossing}, shows a reasonable description to the curves.
Therefore, the singular part is mainly responsible to the behavior of crossing point as a function of $x$ and in the absence of nonsingular part, the approximation in \eq{taucross} can be used as a first time estimate for the crossing points and to fix a corresponding parameter of scale profile function. 

Our numerical results given in this section are purely perturbative results.
Including nonperturbative corrections and hadronization effects are important in precision predictions.
The hadronic effects are power suppressed by $\Lqcd/(\tau Q)$ for $\tau\gg \Lqcd/Q$ in small $\tau$ region
and the correction at the leading power can be parameterized using a nonperturbative parameter $\Omega_1$ \cite{Lee:2006nr,Abbate:2010xh,Mateu:2012nk}. The parameter is universal for any version of 1-jettiness and the dependence on the jet algorithm would remain small in this region. 
One also can take a shape function method that takes into account nonperturbative behavior as well as hadronic power correction \cite{Hoang:2007vb,Kang:2013nha}. 
However more recent studies imply that more careful considerations are required in a scheme related to renormalon subtraction \cite{SCET2021_CLee} and in hadronization effect away from dijet region \cite{Luisoni:2020efy}. Furthermore, recent analysis using Monte Carlo simulations implies its fine-tuning is required to explain HERA measurements \cite{PANIC2021_SHLEE}. Therefore, a quantitative analysis on these effect in 1-jettiness would be beyond the scope of this work and could be done in future project.

%%%%%%%%%%%%%%%%%%%%%%%%%%%%%%%%%%%%%%%%%%%%%%%%%%%%%%
\section{Conclusions}\label{sec:con}
%%%%%%%%%%%%%%%%%%%%%%%%%%%%%%%%%%%%%%%%%%%%%%%%%%%%%%
%%%%%%%%%%%%%%%%%%%%%%%%%%%%%%%%%%%%%%%%%%%%%%%%%%%%%%

In this paper we study the event shape 1-jettiness in DIS at the first order in $\as$ that can be
measured in future experiments. We considered three different jet axes, onto which a particle momentum is projected to compute the value of 1-jettiness.  $\taujt$ is a version with the jettiness axis that is optimally adjusted to minimize the value of 1-jettiness. The other two versions are $\taukt$ and $\tauct$  that take their axes from exclusive jet algorithms such as anti-$k_T$ and the Centauro algorithms in the Breit frame, respectively. We find that $\taujt$ and $\taukt$ are equivalent for the two-body final state, $\ie$ at the order $\as$.

Our main result is the predictions for $\taujt$ and $\tauct$ distributions at the first order in $\as$ analytically expressed in \eqs{AnsBns}{dABns}. They are expressed in the form of the cumulative distribution and in the Appendix the differential distribution is also given. The results are expressed such that one can easily write the structure functions such as $F_1(x,Q^2,\tau)$ and $F_L(x,Q^2,\tau)$ as well as the cross section. The results of $\tauct$ share many terms in common with those of $\taujt$ and their difference depending on $R$ is given so that $\tauct$ is obtained by adding the difference on the top of $\taujt$ result. Comparison to the analytic result for $\taub$ using Breit frame axis are also given in \appx{diff}.

Numerical results of our predictions are presented at different values of $x$ and $Q$. Singular and nonsingular parts are compared in $\taujt$ distributions. For $\tauct$, we studied the difference from $\taujt$ and found that it is sensitive to the value of radius $R$ and the magnitude is in the range of $5\sim 25\%$ of $\taujt$ distribution when the value of $R$ is from $1$ to $1.8$. We also studied the singular-nonsingular crossing point $\tau_\text{cross}$ as a function of $x$ and $Q$. The crossing implies a boundary between resummation and fixed-order regions and the value can be used as a reference point for turning off the resummation. The value of $\tau_\text{cross}$ hence, the resummation region increases with the decreasing value of $x$, and the region gets close to $\tau_\text{max}=1/2$ at $x=0.01$, while the value is less sensitive to $Q$. We found that the $x$ dependence is well explained by the singular part and obtained an approximate expression for $\tau_\text{cross}$, which can be used in the absence of nonsingular part.

Our results provide an important piece of information toward precision predictions of
event shapes in DIS. Our prediction for nonsingular part combined with resummed singular
part can be measured in the future EIC and $\EICC$
and can be used to determine the strong coupling constant and a universal hadronic nonpertubative parameter.

%%%%%%%%%%%%%%%%%%%%%%%%%%%%%%%%%%%%%%%%%%%%%%%%%%%%%%
\begin{acknowledgments}
D.K. likes to thank to Iain Stewart and Christopher Lee for fruitful discussions on the singular and nonsingular crossing and feedback. The work of Z.C. and J.C. are supported in part by the Guangdong Major Project of Basic and Applied Basic Research No. 2020B0301030008. The work of D.K., J.-H.E., and Y.L. is supported by the National Key Research and Development Program of China under Contracts No.~2020YFA0406301 and by the National Natural Science Foundation of China (NSFC) through Grant Nos. 12150610461 and 11875112. The work of J.-H.E. is supported by the NSFC through Grant No. 12105051. The work of J.C. is also supported by the NSFC through Grant Nos. 12025501, 11890714, 12147114, the Strategic Priority Research Program of Chinese Academy of Science under Grant No. XDB34030000
\end{acknowledgments}
%%%%%%%%%%%%%%%%%%%%%%%%%%%%%%%%%%%%%%%%%%%%%%%%%%%%%%

%%%%%%%%%%%%%%%%%%%%%%%%%%%%%%%%%%%%%%%%%%%%%%%%%%%%%%
\appendix
\section*{APPENDIX}
%%%%%%%%%%%%%%%%%%%%%%%%%%%%%%%%%%%%%%%%%%%%%%%%%%%%%%

%%%%%%%%%%%%%%%%%%%%%%%%%%%%%%%%%%%%%%%%%%%%%%%%%%%%%%
\section{Details of calculations}\label{app:detail}

In this section we calculate 1-jettiness QCD cross section.
In derivations of the cross section, we follow the steps done in \cite{Kang:2013nha}. 
The cross section is expressed in the product of lepton and hadronic tensors. 
%%%
\be \label{eq:sigmaLW}
\frac{d\sigma}{dx\,dQ^2\,d\tau} =
 L^{II'}_{\mu\nu}(x,Q^2)W^{II'\mu\nu}(x,Q^2,\tau)\,,
\ee
%%%
where $L^{II'}_{\mu\nu}$ is the lepton tensor given in \cite{Kang:2013nha}
and the index $I={V,A}$ implies vector and axial currents. Although, we consider $II'=VV$
in this paper but we keep these index to maintain generality for a moment.
For vector currents, we have
%%%
\be \label{eq:Lvv}
L^{VV}_{\mu\nu}(x,Q^2)
=-\frac{\alpha^2}{2x^2 s^2}g_{\mu\nu}^T
\,,
\ee
%%%
where $-g_{\mu\nu}^T=-g_{\mu\nu}+2/Q^2(k^\mu k'^\nu +k'^\mu k^\nu )$,
with $k^\mu$ and $k'^\mu$ being the momenta of incoming 
and outgoing leptons and the photon momentum is given by 
$q^\mu = k^\mu - k'^\mu$.
One can show that the lepton tensor is transverse to virtual photon as implied by the Ward identity
$q^\mu L^{VV}_{\mu\nu}\propto q^\mu g_{\mu\nu}^T=0$.
The hadronic tensor $W^{\mu\nu}$ that measures the 1-jettiness from final hadronic state $X$ can be expressed as
%%%
\begin{align}
\label{eq:Wtensor}
 W_{II'}^{\mu\nu}(x,Q^2,\tau) &=
   \sum_X \bra{P}J_{I}^{\mu \dag}(x) \ket{X}\bra{X}J_{I'}^\nu(x)\ket{P}
 (2\pi)^4\delta^4(P + q - p_X)  \delta(\tau-\tau(X))
\,,\nn\\
&=  \sum_{n=1} \int \frac{d\Phi_n}{d\tau}
\bra{P}J_{I}^{\mu \dag}(x) \ket{p_1,\cdots,p_n}\bra{p_1,\cdots,p_n}J_{I'}^\nu(0)\ket{P}
\,,
\end{align}
%%%d
where there is an average over incoming spins implicit. In the second line we specify the number of final particles and the corresponding phase-space integral.

The hadronic tensor is decomposed into several tensors 
%%%
\begin{align}
\label{eq:WFi}
 W_{II'}^{\mu\nu}(x,Q^2,\tau)
&= (4\pi)\bigg[ T^{\mu\nu}_1 F_1(x,Q^2,\tau)+ T^{\mu\nu}_2 \frac{F_2(x,Q^2,\tau)}{P\cdot q}
     +T^{\mu\nu}_3 \frac{F_3(x,Q^2,\tau)}{2 P\cdot q} \bigg]
\,.
\end{align}
%%%
where $F_i$ are called structure functions, which is differential in $\tau$ here and integrating $F_i$ over $\tau$ gives ordinary structure function. The tensor $T^{\mu\nu}_i$ are given by
%%%
\begin{align}
\label{eq:Tmunu}
T_{1\,\mu\nu}&= -g_{\mu\nu}+\frac{q_\mu q_\nu}{q^2}
\,,\nn\\
T_{2\,\mu\nu}&= \bigg(P_\mu-q_\mu \frac{P\cdot q}{q^2}\bigg) \bigg(P_\nu-q_\nu \frac{P\cdot q}{q^2}\bigg)
\,,\nn\\
T_{3\,\mu\nu}&= -i \e_{\mu\nu\alpha \beta}q^\alpha P^\beta
 \,.
\end{align}
%%%
Multiplying $g_{\mu\nu}$ and $P_\mu P_\nu$ by \eq{Wtensor}, one obtains linear combinations of $F_1$ and $F_2$.
Solving for $F_1$ and $F_2$, they are expressed in terms of the hadronic tensor.
Similarly, $F_3$ can be obtained by multiplying $T_{3\,\mu\nu}$ by \eq{Wtensor}.
%%%
\begin{align}
\label{eq:Fi}
F_1(x,Q^2,\tau)
&=\frac{1}{8\pi (1-\e)} \bigg(-g_{\mu\nu}+\frac{4x^2}{Q^2}P_\mu P_\nu \bigg) W^{\mu\nu}_{II'}
\,,\nn\\
F_2(x,Q^2,\tau)&=\frac{x}{4\pi (1-\e)} \bigg(-g_{\mu\nu}+(3-2\e)\frac{4x^2}{Q^2}P_\mu P_\nu\bigg) W^{\mu\nu}_{II'}
\,,\nn\\
F_L(x,Q^2,\tau)&=\frac{2x^3}{\pi Q^2}\frac{-q^2}{(P\cdot q)^2} P_\mu P_\nu W^{\mu\nu}_{II'}
=F_2-2x\, F_1
\,,\nn\\
F_3(x,Q^2,\tau)
&=\frac{x}{2\pi(1-\e)(1-2\e)}
\frac{ q^\alpha P^\beta \e_{\alpha \beta \mu\nu}}{Q^2} W^{\mu\nu}_{II'}
 \,.
\end{align}
%%%
In this paper we consider $II'=VV$, there is no $F_3$ contribution. From hereafter we drop the index for the hadronic tensor $W^{\mu\nu}=W^{\mu\nu}_{VV}$ and one needs to calculate $g_{\mu\nu}W^{\mu\nu}$, $P_\mu P_\nu W^{\mu\nu}$.
The cross section in \eq{sigmaLW} in terms of the structure functions $F_i$ is given by \eq{sigma-Fi}.

%%%%%%%%%%%%%%%%%%%%%%%%%%%%%%%%%%%%%%%%%%%%%%%%%%%%%%%%%%%%%%%%%%%%%%%%%%%%%%%%
\subsection{Phase space integral}
\label{sec:phase space}

The phase-space integral can be explicitly written as
%%%
\begin{align}
\label{dPhin}
\int \frac{d\Phi_n}{d\tau}&=
\mu^{2\e (n-1)}\int \prod_{i=1}^{n} \frac{d^d p_i}{(2\pi)^d} (2\pi) \delta(p_i^2)
\, (2\pi)^d\delta^d(P + q - \sum_i p_i)  \delta(\tau-\tau(\{p_i\}))
\,,
\end{align}
%%%
where $d=4-2\e$.
For $n=1$,
%%%
\begin{align}
\label{eq:dPhi1}
\int \frac{d\Phi_1}{d\tau}&=
\frac{2\pi}{Q^2}\delta(1-z)\delta(\tau)
\,.
\end{align}
%%%

For $n=2$,
%%%
\begin{align}
\label{eq:dPhi2-v}
\int \frac{d\Phi_2}{d\tau}
&=
\frac{1}{8\pi Q} \frac{(4\pi\mu^2)^\e}{ \Ga(1-\e)}
\,\int \frac{dp_2^- }{(p_2^+ p_2^-)^\e} \delta\big(\tau-\tau(p_1,p_2)\big)
\nn\\
&=
M(\e)\,
\,\int dv\frac{\theta(v)\,\theta(1-v) }{v^\e (1-v)^\e} \delta\big(\tau-\tau(v,z)\big)
\,,\\\nn
M(\e)
&=
\frac{1}{8\pi}\frac{(4\pi \mu^2/Q^2)^\e}{\Ga(1-\e)}\bigg(\frac{z}{1-z}\bigg)^\e
\,.
\end{align}
%%%

\fig{vz-space} shows three and four regions in $v$-$z$ space divided for $\taujtkt$ and for $\tauct$, respectively.
Then the phase-space integral in \eq{dPhi2-v} splits into three pieces as
%%%
\begin{subequations}\label{eq:dPhi2}
\begin{align}
\int_{\text{I}} \frac{d \Phi_2}{d\taujt}
&=
M(\e)\,\Theta_0(\tau,z)
\frac{1}{(1-\tau)^\e \tau^\e}\int dv\,\delta(v-1+\tau)\,,
\label{eq:dPhi2-I} 
\\
\int_{\text{II}} \frac{d \Phi_2}{d\taujt}
&=
M(\e)\,\Theta_0(\tau,z)
\frac{1}{(1-\tau)^\e \tau^\e}\int dv\,\delta(v-\tau)\,,
\label{eq:dPhi2-II} 
\\
\int_{\text{III}} \frac{d \Phi_2}{d\taujt}
&=
M(\e)\theta\left(z-\frac{2}{3}\right)\delta\left(\tau-\frac{1-z}{z}\right)
\int^{1-\tau}_{\tau}\frac{dv}{v^\e (1-v)^\e}\,,
\label{eq:dPhi2-III} \\
\int_{\text{IV}}\left( \frac{d \Phi_2}{d\tauct}- \frac{d \Phi_2}{d\taujt}\right)
&=M(0)\left[\delta\left(\tau -\frac{(2z-1)(1-z)}{z^2}\right)-\delta\left(\tau-\frac{1-z}{z}\right) \right]\int^{v_+(z,R)}_{v_-(z,R)} dv
\,,
\label{eq:dPhi2-IV} 
\end{align}
\end{subequations}
%%%
where $\Theta_0$ and $v_\pm$ are given in \eqs{theta0}{vpm-zc}, respectively.
Note that the region III in \eq{dPhi2-III} is that of  $\taujt$ shown in \fig{vz-space}. In $\tauct$ computation we use the same expression, which incorrectly include the region IV then, in \eq{dPhi2-IV} the incorrect contribution from $\taujt$ is subtracted.  It makes \eq{dPhi2-IV} finite so that we can set $\e$ to zero and also makes the expression of $\tauct$ cross section simple with an additional term added to $\taujt$ cross section.

%%%%%%%%%%%%%%%%%%%%%%%%%%%%%%%%%%%%%%%%%%%%%%%%%%%%%%%%%%%%%%%%%%%%%%%%%%%%%%%%
\subsection{Hadronic tensor for incoming quark}
\label{ssec:htensor-q}

The hadronic tensor defined in \eq{Wtensor}, the matrix element with incoming proton can be factorized into convolutions of proton PDF and Wilson coefficients.  The Wilson coefficients can be determined by matching the factorization formula with incoming
parton to the hadronic tensor for the parton state.  The hadronic
tensors $W^q_{\mu\nu}$ and $W^g_{\mu\nu}$ for initial quark and gluon,
respectively should be calculated. To $\cO(\as)$ $W^q_{\mu\nu}$
involves one-loop calculation and $W^g_{\mu\nu}$ is simply tree level
calculation.
In this section, we calculate $W^q_{\mu\nu}$.
$W^q_{\mu\nu}$ receives contributions from tree, virtual, and real diagrams so can be written as
%%%
\begin{align}
\label{eq:W-decompose}
 W^{q}_{\mu\nu}=W_{\mu\nu}^{(0)}+W_{\mu\nu}^\text{vir}+W_{\mu\nu}^\text{real}
\,.
\end{align}
%%%
Here, we suppressed the superscript $q$ on right side and we will suppress it in the middle
of calculations.

%%%%%%%%%%%%%%%%%%%%%%%%%%%%%%%%%%%%%%%%%%%%%%%%%%%%%%%%%%%%%%%%%%
\subsubsection{Tree-level and virtual contributions}
The tree-level amplitude is
$\cM_\mu^{(0)}=Q_f\bar{u}(p_1)\gamma_\mu u(P)$.
The tree-level hadronic tensor is given by
%%%
\begin{align} %\label{eq:Wtree}
W_{\mu\nu}^{(0)}
&= \frac{1}{2}\sum_{\sigma} \int \frac{d\Phi_1}{d\tau}\, \cM^{(0)}_\mu \cM^{(0)\,*}_\nu
= -2  \pi Q_f^2g^T_{\mu\nu}\delta(1-x)\delta(\tau) 
\nn \,,
\end{align}
%%%
where we performed the spin average explicitly over quark spins $\sigma$
then used the 1-body phase space in \eq{dPhi1}. 
The transverse metric is $g^T_{\mu\nu} = g_{\mu\nu} -(n_{z\mu}\bn_{z\nu} + n_{z\nu}\bn_{z\mu}) /2$. This gives the projected hadronic tensors
%%%
\begin{align} \label{eq:gWtree}
-g^{\mu\nu}  W_{\mu\nu}^{(0)}   &= 4\pi Q_f^2 \delta (1-z) \delta (\tau) \,,\\
P^\mu P^\nu\,W_{\mu\nu}^{(0)} &= 0
\,,
\end{align}
%%%
where the second line vanishes by the Dirac equation of massless particle $P\!\!\!\!\slash\, u(P)=0$. 

The virtual contribution can be found from many literature for instance (14.19) \cite{Sterman:1994ce} and the contribution is the same as that of $\taub$ in \cite{Kang:2014qba}
%%%
\begin{align} \label{eq:gW-vir}
-g^{\mu\nu}W_{\mu\nu}^\text{vir}
&= -4\as Q_f^2 C_F (1-\e) \bigg(\frac{4\pi\mu^2}{Q^2}\bigg)^\e
\frac{\Ga(1+\e) \Ga(1-\e)^2}{\Ga(1-2\e)}
\bigg(\frac{1}{\e^2} +\frac{3}{2\e}+4\bigg)
\delta(1-z)\delta(\tau)
\,\nn\\
&=
2\as Q_f^2 C_F (1-\e)
\bigg[ -\frac{2}{\e^2}-\frac{1}{\e}\bigg(2L+3 \bigg)
-L^2-3L+\frac{\pi^2}{6}-8
\bigg]
\delta(1-z)\delta(\tau)
\,,\nn\\
P^\mu P^\nu\,W_{\mu\nu}^\text{vir} &= 0
\,,
\end{align}
%%%
where $L=\ln\frac{\mu^2}{Q^2}$.
The factor $(1-\e)$ is not expanded because it is to cancel $1/(1-\e)$ in \eq{Fi}.
In the second line of \eq{gW-vir} we have used the $\MSbar$ scheme by re-scaling the scale $\mu^2$ by $e^{\gamma_E}/(4\pi)$ such that we use following replacement
%%%
\begin{align}
\label{eq:mu-rescal}
\frac{(4\pi\mu^2)^\e}{\Ga(1-\e)}
\to
\frac{(\mu^2 e^{\gamma_E})^\e}{\Ga(1-\e)}
=\mu^{2\e} \bigg( 1-\e^2\frac{\pi^2}{12}+\cO(\e^3)\bigg)
\,.
\end{align}
%%%
Note that the finite part in \eq{gW-vir} is the $\as$ term in the hard function in \cite{Kang:2013nha}, derived in SCET in \cite{Bauer:2003di,Manohar:2003vb}.
The $P^\mu P^\nu\,W_{\mu\nu}^\text{vir}$ is zero because $P\!\!\!\!\slash\, u(P)=0$.

%%%%%%%%%%%%%%%%%%%%%%%%%%%%%%%%%%%%%%%%%%%%%%%%%%%%%%%%%%%%%%%%%%%%%%%%%%%%%%%%
\subsubsection{Real contribution}\label{sssec:real}

The definition of $\tau$ in \eqs{taujt}{tauct} divide the phase space into three or four
regions as shown in \fig{vz-space} hence the 2-body phase-space is accordingly divided as follows
%%%
\begin{align}
\label{W-abcd}
 W_{\mu\nu}^\text{real}&=\sum_{i=\text{I}}^\text{IV} W_{\mu\nu}^{(i)}
\,,\\
\label{W-MM}
 W_{\mu\nu}^{(i)}
&= \frac{1}{2} \sum_{\sigma}\int_{(i)}\frac{d\Phi_2}{d\tau}
\,\,\cM_\mu^{\text{real}} \cM_\nu^{\text{real}\,*}
\hspace{1cm}i\in\{\text{I},\text{II},\text{III},\text{IV}\}
\,,
\end{align}
%%%
where in the last line the factor $1/2$ takes account the average over quark spins $\sigma$.
In this subsection, we discuss the results of regions I,II, and III,
which is relevant to $\taujt$ and for the region IV, which is IR finite and simpler, we just give the final result in \appx{tauct}.

The real emission contribution to the two projections of the color-averaged squared amplitudes that we need are given by
%%%
\begin{align}
\label{eq:gMM-real}
-g^{\mu\nu}\cM_\mu^{\text{real}} \cM_\nu^{\text{real}\,*}
&= 32\pi\as C_F Q_f^2  (1-\e)
\bigg[
 (1-\e) \bigg( \frac{1-z}{v}
+ \frac{v}{1-z} \bigg)
+2 \frac{z}{1-z}\frac{1-v}{v}
+2\e
\bigg]
\,,\\
\label{eq:ppM-real}
P^\mu P^\nu\,\cM_\mu^{\text{real}} \cM_\nu^{\text{real}\,*} &=
16\pi\as  C_F Q_f^2 Q^2(1-\e) \frac{1-v}{z}
\,,
\end{align}
%%%
where the first line can be found from (14.21) in \cite{Sterman:1994ce}.

For $P^\mu P^\nu\, W_{\mu\nu}^{(i)}$, there is no singular term and
we can safely set $\e =0$. Including the spin average, we obtain
%%%
\begin{subequations}\label{eq:ppW}
\begin{align}
\label{eq:ppW-I}
P^\mu P^\nu\, W_{\mu\nu}^{\text{I}}
&=\as C_F  Q_f^2Q^2\,\frac{\tau}{z}  \Theta_0(\tau,z)
\,,\\\vspace{5pts}
\label{eq:ppW-II}
P^\mu P^\nu\, W_{\mu\nu}^{\text{II}}
&=
\as C_F Q_f^2 Q^2\,\frac{1-\tau}{z}
\, \Theta_0(\tau,z)
\,,\\\vspace{5pts}
\label{eq:ppW-III}
P^\mu P^\nu\, W_{\mu\nu}^{\text{III}}
&= \as C_F Q_f^2 Q^2\,\theta(z-2/3)\delta\left(\tau-\frac{1-z}{z}\right)
\frac{(1-2\tau)(1+\tau)}{2}
\,,
\end{align}
\end{subequations}
%%%
where the $\Theta_0(\tau,z)$ is given in \eq{theta0}. The sum of \eq{ppW} is given by
%%%
\begin{align}\label{eq:ppW-q}
P^\mu P^\nu\, W^q_{\mu\nu}=\as C_F  Q_f^2 Q^2
\,\Theta_0(\tau,z)
\left[
\frac{1}{z}
+\theta\left(z-\frac{2}{3}\right)\delta\left(\tau-\frac{1-z}{z}\right)
\frac{\left(1-2\tau\right)\left(1+\tau\right)}{2}
\right]
\,,
\end{align}
%%%
where the superscript $q$ representing the incoming quark is made explicit, again.
Note that the tree-level and virtual contributions are zero and the real contribution is the total.
%%%
\begin{subequations}\label{eq:gW}
  \begin{align}
    -g^{\mu\nu}W^{\text{(I)}}_{\mu\nu} &=
  16\pi\alpha_s C_F Q^2_f M(\e)\Theta_0(\tau,z)
  (1-\tau)^{-\e} \tau^{-\e}(1-\e)
  \nn\\
  &\quad\times
  \,\left[(1-\e)\left(\frac{1-z}{1-\tau}+\frac{1-\tau}{1-z}\right)
  +2\frac{z}{1-z}\frac{\tau}{1-\tau}+2\e \right],
  \\
  -g^{\mu\nu}W^{{\textrm{(II)}}}&=
  16\pi\alpha_s C_F Q^2_f M(\e)\Theta_0(\tau,z)
  (1-\tau)^{-\e} \tau^{-\e}(1-\e)
   \nn\\
  &\quad\times
  \,\left[(1-\e)\left(\frac{1-z}{\tau}+\frac{\tau}{1-z}\right)
  +2\frac{z}{1-z}\frac{1-\tau}{\tau}+2\e\right],
  \\
  -g^{\mu\nu}W^{{\textrm{(III)}}}&=
  16\pi\alpha_s C_F Q^2_f M(\e)\theta\left(z-\frac{2}{3}\right)
  \delta\left(\tau-\frac{1-z}{z}\right)(1-\e)\nn
  \\&\quad
  \times\int^{1-\tau}_{\tau}\frac{dv}{v^\e (1-v)^\e}
  \left[(1-\e)\left(\frac{1-z}{v}+\frac{v}{1-z}\right)+2\frac{z}{1-z}\frac{1-v}{v}+2\e \right].
  \end{align}
\end{subequations}
%%%
The regions $\textrm{I}$ and $\textrm{II}$ share the same $\Theta_0(\tau,z)$, so we combine them and expand in $\e$
%%%
\begin{align}\label{eq:gW-I-II}
  & -g^{\mu\nu}W^{\textrm{(I)}+\textrm{(II)}}_{\mu\nu}
 =
  2\as C_F Q^2_f \left(\frac{\mu^2}{Q^2}\right)^\e
  \frac{e^{\e\gamma_E }}{\Ga(1-\e)}z^\e (1-\e)
  \nn\\
  & \quad\times
  \left[\theta\left(-\tau+\frac{1-z}{z} \right)\theta\left(z-\frac{2}{3} \right)
          +\theta\left(-\tau+\frac{1}{2} \right)\theta\left(-z+\frac{2}{3} \right)\right]
  \nonumber\\
  &\quad\times
\left\{
	(1-\e)\left(
 	(1-z)^{1-\e} \frac{1}{\tau^{1+\e}} \frac{1}{(1-\tau)^{1+\e}} 
  + \left(1-\tau\right)^{-\e} \tau^{-\e} \frac{1}{(1-z)^{1+\e}}  
  \right)
  \right.
   \nonumber\\    
  & \quad\quad
   +2z\tau^{1-\e}
 {\frac{1}{(1-z)^{1+\e}}}\frac{1}{(1-\tau)^{1+\e}}
  \nn\\
  &\quad\quad
  \left.
  +2z(1-\tau)^{1-\e}{\frac{1}{(1-z)^{1+\e}}}
 {{\frac{1}{\tau^{1+\e}}}}
  +4\e (1-\tau)^{-\e} \tau^{-\e} \frac{1}{(1-z)^{\e}} 
  \right\},
\end{align}
%%%
where $\Theta_0(\tau,z)$ was divided  into two parts, and the first part multiplied by terms $1/\tau^{1+\e}$ and $1/(1-z)^{1+\e}$ gives $1/\e$ pole, while with the other part the only term like $1/\tau^{1+\e}$ gives the pole.
We now expand above equation in powers of $\e$. For the term like 
$ \theta(-\tau+\tfrac{z}{1-z})/\tau^{1+\e}/(1-z)^{1+\e}$ we use
 a plus distribution identity in Appendix~A of \cite{Kang:2014qba}.
%%%
\begin{align}\label{eq:gW-I-II-f}
-g^{\mu\nu}W^{\textrm{(I+II)}}_{\mu\nu}
&=2\as C_F Q^2_f \left(\frac{\mu^2}{Q^2}\right)^\e (1-\e)
\Theta_0(\tau,z)
\nn\\
&\quad \times
\, \left[\left(\frac{1}{\e^2}+\frac{3}{2\e}\right)
\delta(\tau)\delta(1-z)-\frac{P_{\text{qq}}(z)}{\e}\delta(\tau)
+I^s_{\textrm{fin}}+I^{ns}_{\textrm{fin}}
 \right],
\end{align}
%%%
where $P_{qq}$ is the splitting function in \eq{Pqqqg}.
The functions $I^s_{\textrm{fin}}$ and $I^{ns}_{\textrm{fin}}$ contribute to singular and nonsingular parts, respectively.
%%%
\begin{align}\label{eq:Ifin-s}
  I^s_{\textrm{fin}}&=\delta(\tau)
  \left[(1+z^2)\cL_{1}(1-z)-\frac{\pi^2}{12}\delta(1-z)
  +1-z-(1+z^2)\ln z~\cL_0 (1-z)   \right]
  \nonumber\\ & \quad
  +\cL_0(\tau)\left[P_{\mathrm{qq}}(z)-\frac{3}{2}\delta(1-z) \right]
  -2\cL_1 (\tau)\delta(1-z),\nonumber \\
  I^{ns}_{\textrm{fin}}&=
  \frac{(3\tau-1)\cL_0 (1-z)-4\tau-z+3}{1-\tau}+2z~I_{ns}(\tau,1-z)
\,,
\end{align}
%%%
where $I_{ns}$ is identical to corresponding part in $\taub$ and is given in Eq.~(A.8) of \cite{Kang:2014qba}.
We use following relations to compress \eq{gW-I-II-f}
%%%
\begin{align}\label{eq:Pqq}
&P_{qq}(z)=\bigg[\frac{1+z^2}{1-z}\bigg]_+
= 2 \cL_0(1-z) +\frac32\delta(1-z)-(1+z)=(1+z^2)\cL_0(1-z)+\frac32 \delta(1-z)
\,,\nn\\
&(1+z^2)\cL_0(1-z)= 2\cL_0(1-z)-(1+z)
\,,\nn\\
&(1+z^2)\cL_1(1-z)=2\cL_1(1-z)-(1+z)\ln(1-z)
\,,
\end{align}
%%%
where standard plus distributions $\cL_n(z)$ are defined by \cite{Kang:2014qba}
%%%
\begin{align}\label{eq:plus-def}
    &\cL_n(z)\equiv\lim_{\varepsilon \to 0}\frac{d}{d z}\bigg[\frac{\theta(z-\varepsilon)\ln^{n+1}z}{n+1}\bigg]=\bigg[\frac{\theta(z)\ln^n(z)}{z}\bigg]_+
\,.
\end{align}
%%%
\,With a test function $g(z)$ well behaving near $z=0$, the integration against the function gives
%%%
\begin{align}
    &\int^z_0 d z^{'}\cL_n(z^{'})g(z^{'})=\int^z_0 d z^{'}\frac{\ln^n z^{'}}{z^{'}}[g(z^{'})-g(0)]+g(0)\frac{\ln^{n+1} z}{n+1}
\,.
\end{align}
%%%
A variation of $\cL_n$ with a variable lower bound $z_0$ and its integral is given by
%%%
\begin{align}
\cL_n(z,z_0)&=\lim_{\varepsilon \to 0}\frac{d}{d z}\bigg[\frac{\theta(z-z_0-\varepsilon)\ln^{n+1}z}{n+1}\bigg]
\,,\nn\\
\int^z_0 d z^{'}\cL_n(z^{'},z_0)g(z^{'})&=\int^z_{z_0} d z^{'}\frac{\ln^n z^{'}}{z^{'}}[g(z^{'})-g(z_0)]+g(z_0)\frac{\ln^{n+1} z}{n+1}
\,.
\end{align}
%%%%

For the contribution from the region III, it contains singular terms when $\tau$ or, $1-z$ approaches zero. We isolate the singular part carefully by subtracting and adding singular terms 
%%%
\begin{subequations}\label{eq:gW-III-0123}
\begin{align}
  -g^{\mu\nu}W^{{\textrm{(III)}}}_{\mu\nu}&= 
  2\as C_F Q^2_f
  \left(\frac{\mu^2}{Q^2}\right)^\e \frac{e^{\e \gamma_{E} }}{\Ga(1-\e)}\theta\left(z-\frac{2}{3}\right)
  \delta \left(\tau-\frac{1-z}{z}\right)(1-\e)
  \nn\\
  &\quad \times
  \left[ (1-\e)(I_1+I_2 )+2I_3 \right],
  \label{eq:gW-III-0}
  \\\vspace{5pts}
  I_1&= \frac{\tau^{1-\e}}{1+\tau}\int^{1-\tau}_{\tau}
  dv\frac{1}{v^{1+\e}(1-v)^\e}=\frac{\tau}{1+\tau}
  \ln\bigg(\frac{1-\tau}{\tau}\bigg),
  \label{eq:gW-III-1}\\\vspace{5pts}
  I_2&=\frac{1+\tau}{\tau^{1+\e}}\int^{1-\tau}_{\tau}
  dv\frac{v^{1-\e}}{(1-v)^{\e}}
  \nn\\
  &=
  \frac{1+\tau}{\tau^{1+\e}} \left[\frac{1}{2}-\tau+\e\big(1-2\tau)+\e(\tau-1)\ln(1-\tau)+\e\tau\ln\tau
  \big) \right],\
  \label{eq:gW-III-2}\\\vspace{5pts}
  I_3&=\frac{1}{\tau^{1+\e}}\int^{1-\tau}_{\tau}dv\,
  \frac{(1-v)^{1-\e}}{v^{1+\e}}
  \nn\\
  &=
  \frac{1}{\tau^{1+\e}}\int^{1-\tau}_{\tau}dv
  \frac{(1-v)^{1-\e}-(1-\tau)^{1-\e}}{v^{1+\e}}
  +\frac{1}{\e}\left[\frac{(1-\tau)^{1-\e}}{\tau^{1+2\e}} -\frac{(1-\tau)^{1-2\e}}{\tau^{1+\e}}\right]
  \,,
   \label{eq:gW-3}
\end{align}
\end{subequations}
%%%
where we used the delta function in the first line and replace $z$ by $\tau$. Finally we expand \eq{gW-III-0} in $\e$ 
%%%
\begin{align}
-g^{\mu\nu}W^{(\text{III})}_{\mu\nu}=&2\as C_FQ_f \left(\frac{\mu^2}{Q^2}\right)^\e (1-\e)\,\theta\left(z-\frac{3}{2}\right)
\delta\left(\tau-\onez\right)
\nn\\
 &\times
\left[ \delta(\tau)\big(\frac{1}{\e^2}+\frac{3}{2\e}+\frac{7}{2}-\frac{5\pi^2}{12}\big)
-\frac{3}{2}\cL_0(\tau)-2\cL_1(\tau)+\frac{7}{2}
\right.
\nn\\
&\qquad\qquad \left. +
\frac{\tau^2+2\tau+2}{\tau(1+\tau)}\ln(1-\tau)-\frac{\tau}{1+\tau}\ln\tau-\tau
\right]
\,.
\label{eq:gW-III-f}
\end{align}
%%%
It is a non-trivial cross check to show that
sum of $-g^{\mu\nu}W_{\mu \nu}^{(i)}(\tau)$ and $P^\mu P^\nu W_{\mu \nu}^{(i)}(\tau)$ over $i$ and its integration over $\tau$ are equivalent to the inclusive results that simply obtained by integrating the phase space measures in \eqss{dPhi2-I}{dPhi2-II}{dPhi2-III} over $\tau$ then by carrying out integral against amplitudes \eqs{gMM-real}{ppM-real} over $v$.

Now we collect all the pieces and put them together.
All IR divergences with $1/\e^2$ and $1/\e$ are cancelled when virtual
part in \eq{gW-vir} and real parts in \eqss{gWtree}{gW-I-II-f}{gW-III-f} are combined except for the IR divergence associated with 1-loop quark PDF. Let us first highlight a few terms
%%%
\begin{align}
-g^{\mu\nu}\, W_{\mu\nu}
&\propto\bigg[
 \delta(1-z)
\big(L^2+3L\big)-\frac{P_{qq}(z)}{\e}- P_{qq}(z) L
\bigg]
\nn\\
&\qquad
+\text{terms from finite part}
\label{eq:gWdiv}
\,,
\end{align}
%%%
where $L=\ln \frac{\mu^2}{Q^2}$.
First, the two logarithmic terms above are cancelled by the same terms in virtual part \eq{gW-vir}.
The $1/\e$ term above will be replaced by 1-loop correction of the proton PDF during matching procedure.
The logarithmic scale dependence proportional to $P_{qq}$ in the last term should cancel the same scale dependence from RG evolution of the proton PDF. Therefore, all $\mu$ dependence at $\cO(\as)$ are cancelled. 

Finally we have
%%%
\begin{align}
-g^{\mu\nu}\, W^q_{\mu\nu}(z,\tau,Q^2)
&=4\pi  Q_f^2 \delta(1-z) \delta(\tau)
\nn\\&\quad
+2\as Q_f^2 C_F (1-\e)\Theta_0(\tau,z)
\bigg\{
\delta(\tau) \,\left[-\frac{ P_{qq}(z)}{\e} +S_{-1}^q(z)\right]
+\cL_0(\tau) \,S_{0}^q(\tau,z)
\nn\\&\quad\quad
+\cL_1(\tau)\,S_{1}^q(\tau,z)
+R^q(\tau,z)
+\delta\left(\tau-\onez\right)\,\Delta^q(\tau)
\bigg\}
\label{eq:gW-q}
\,,
\end{align}
%%%
where the superscript $q$ denotes the contribution from the incoming quark.
The functions $S,~R,~\Delta$ are given by
%%%
\begin{align}
S_{-1}^q(z)&=
-P_{qq}(z) \ln \frac{\mu^2}{Q^2} 
+ (1+z^2)\cL_1(1-z)
-\bigg( \frac{9}{2}+\frac{\pi^2}{3}\bigg)\delta(1-z) +1-z
-\Pqqr \ln z
\,,\nn\\
S_{0}^q(\tau,z)&=
2z\,\cL_0\left(1-z,\tfrac{\tau}{1+\tau}\right) -\frac{3}{2}\delta\left(\tau-\onez\right)+(1-z)
\,,\nn\\
S_{1}^q(\tau,z)&=
-2(2+\tau)\,\delta \left(\tau-\onez\right)
\,,\nn\\
R^q(\tau,z)&=
\frac{1-4z}{1-z}+\Pqqr\frac{1}{1-\tau}
\,,\nn\\
\Delta^q(\tau)&=
\frac{\tau^2+2\tau+2}{\tau(1+\tau)}\ln (1-\tau)-\frac{\tau}{1+\tau}\ln \tau+\frac{2(1+\tau)}{\tau}\ln (1+\tau)-\tau+\frac{7}{2}
\label{eq:coeffq}
\,.
\end{align}
%%%
The functions $S^q_n$ are similar to those in $\taub$. $S^q_{-1}$ is the same except for the last $\ln z$ term and in $S^q_{0,1}$ the delta function terms contains additional factors $1+\tau$ associated with change of variable in the delta function $\delta(\tau-\frac{1-z}{z})=(1+\tau)^{-2}\delta(\tfrac{1}{1+\tau}-z)$.

%%%%%%%%%%%%%%%%%%%%%%%%%%%%%%%%%%%%%%%%%%%%%%%%%%%%%%%%%%%%%%%%%%%%%%%%%%%%%%%%
\subsection{Hadronic tensor for incoming gluon}  \label{ssec:tensor-g}
A process with the initial gluon $g+\gamma^*\to q\bar{q}$ starts at $\cO(\as)$ and this process is the tree level.
The tree level amplitude averaged over incoming colors is given by 
%%%
\begin{align} \label{eq:gMM-g}
-g^{\mu\nu}\cM_\mu^{g} \cM_\nu^{g\,*}
&= 32\pi\as T_F \sum_f Q_f^2  (1-\e)
\bigg[
 (1-\e) \bigg( \frac{1-v}{v}
+ \frac{v}{1-v} \bigg)
-2 \frac{z(1-z)}{v(1-v)}
-2\e
\bigg]
\,,\\
\label{eq:ppM-g}
P^\mu P^\nu\,\cM_\mu^{g} \cM_\nu^{g\,*} &=
32\pi\as  T_F \sum_f Q_f^2 \, Q^2(1-\e) \frac{1-z}{z}
\,,
\end{align}
%%%
where the sums go over the all flavors $f\in \{u,d,s,c,b\}$ allowed by the energy.
Note the $(1-\e)$ factors in \eqs{gMM-g}{ppM-g} are going to be canceled by the same factor in the structure functions \eq{Fi}.
Integrating \eq{ppM-g} over the two-body final-state phase space,
%%%
\be
W_{\mu\nu}^{g} = \frac{1}{2-2\e} \sum_{\lambda} \int \frac {d\Phi_2}{d\tau} \cM_{\mu}^{g} \cM_\nu^{g*}\,,
\ee
%%%
where the superscript $g$ representing the incoming gluon  and the prefactor $(2-2\e)$ accounts for the average over incoming gluon polarization $\lambda$ in $D=4-2\e$ dimensions.
Then, we have
%%%
\begin{align}\label{eq:ppW-g}
P^\mu P^\nu\, W^g_{\mu\nu}= 4\as T_F  Q_f^2 Q^2\Theta_0
\,(1-z)
\left[
\frac{1}{z}
+\theta\left(z-\frac23\right)\delta\left(\tau-\frac{1-z}{z}\right)\,
\frac{(1-2\tau)(1+\tau)}{2}
\right]
\,,
\end{align}
%%%
The contributions $-g^{\mu\nu}W_{\mu\nu}^{(i)}$ for $i=\text{I}$ and for $\text{II}$ are identical  and involve singular terms
%%%
\begin{align}\label{eq:ggW-I}
-g^{\mu\nu}\, W_{\mu\nu}^{(\text{I})}&=-g^{\mu\nu}\, W_{\mu\nu}^{(\text{II})}
\nn\\
&=2\as T_F \sum_f Q_f^2  \frac{(\mu^2 e^{\gamma_E}/Q^2)^\e}{\Ga(1-\e)}\bigg( \frac{z}{1-z}\bigg)^\e(1-\e) \Theta_0(\tau,z)
\nn\\
&\quad \times\left[ \bigg(\frac{\tau^{1-\e}}{(1-\tau)^{1+\e}}
+ \frac{(1-\tau)^{1-\e}}{{\tau^{1+\e}}}    \bigg)
-\frac{2}{1-\e}\frac{z(1-z)}{{\tau^{1+\e}}(1-\tau)^{1+\e}}
\right]
\nn\\
&=
2\as T_F \sum_f Q_f^2 \bigg(\frac{\mu^2}{Q^2}\bigg)^\e (1-\e)\,\Theta_0(\tau,z)
\bigg[ -\frac{P_{qg}(z)}{\e}\delta(\tau)
\nn \\
& \quad
+\bigg(1-P_{qg}(z)+P_{qg}(z)\ln\frac{1-z}{z}\bigg)\,\delta(\tau)
+P_{qg}(z)\cL_0(\tau)+
\frac{P_{qg}(z)}{1-\tau}-2 \bigg]
\,,
\end{align}
%%%
where the splitting function $P_{qg}(z)$ is given in \eq{Pqqqg}.

The contraction $-g^{\mu\nu}\, W_{\mu\nu}^{(\text{III})}$ also involves singular terms when $\tau \rightarrow 0$ and $\tau \rightarrow 1$, so doing similar expansions as in \eq{gW-3}, we get
%%%
\begin{align}\label{eq:ggW-III}
  -g^{\mu\nu}\, W_{\mu\nu}^{(\text{III})}
  &=
  2\as T_F \sum_f Q_f^2 
   \frac{(\mu^2 e^{\gamma_E}/Q^2)^\e}{\Ga(1-\e)}\bigg( \frac{z}{1-z}\bigg)^\e\, \deltatz
  \nn \\ & \quad
  \times \int^{1-\tau}_{\tau} dv \left[(1-\e)\bigg({\tfrac{(1-v)^{1-\e}}{v^{1+\e}}}
  +{\tfrac{v^{1-\e}}{(1-v)^{1+\e}}}  \bigg) -2\tfrac{z(1-z)}{{v^{1+\e}}{(1-v)^{1+\e}} } -\tfrac{2\e}{v^\e (1-v)^\e} \right]
  \nn \\ \!\!\!\!\!\! &=
  2\as T_F \sum_f Q_f^2 \left(\tfrac{\mu^2}{Q^2}\right)^\e (1-\e)\,\deltatz\, \left[ 4\tau-2+2P_{qg}(z)\ln \tfrac{1-\tau}{\tau} \right]\,.
\end{align}
%%%

By putting all the  $-g^{\mu\nu}\, W_{\mu\nu}^{(i)}$ in \eqs{ggW-I}{ggW-III} together we have
%%%
\begin{align}
\label{eq:gW-g}
-g^{\mu\nu}\, W_{\mu\nu}^{g}
&= 4\as T_F \sum_f Q_f^2(1-\e)
\Theta_0(\tau,z)
\bigg[-\frac{P_{qg}(z)}{\e}\delta(\tau)
\nn\\
&\qquad 
+ S^g_{-1}\delta(\tau)+S^g_{0}\cL_0(\tau)+R^g(\tau,z)
+\delta\bigg(\tau-\onez \bigg)\Delta^g(\tau,z)
\bigg]
\,,\\
  S^g_{-1}(z)&= -P_{qg}(z)+ 1+P_{qg}(z)\bigg( \ln\frac{1-z}{z}-\ln\frac{\mu^2}{Q^2} \bigg) 
  \,,
  \nn\\  
  S^g_0(\tau,z)&=P_{qg}(z)\,,\nn\\
  R^g(\tau,z)&=\frac{P_{qg}(z)}{1-\tau}-2
   \,,
   \nn\\
  \Delta^g(\tau,z)&=2\tau-1+P_{qg}(z)\ln\frac{1-\tau}{\tau}
  \,.
\end{align}
%%%

%%%%%%%%%%%%%%%%%%%%%%%%%%%%%%%%%%%%%%%%%%%%%%%%%%%%%%%%%%%%%%%%%%%%%%%%%%%%%%%%
\subsection{Nonsingular part and cumulative results}\label{ssec:nonsing}

In this subsection, we give the expressions for the nonsingular terms of the hadronic tensors obtained in previous sections. The tensors reduce to singular terms having $\delta(\tau)$, or $\cL_n(\tau)$
as $\tau\to 0$. In the contributions $P_\mu P_\nu W_{\mu\nu}$ in \eqs{ppW-q}{ppW-g} there is no such terms and they are purely nonsingular. 

In the gluon process $-g^{\mu\nu}W_{\mu\nu}^g$ in \eq{gW-g} contains terms with
$\delta(\tau)$ and $\cL_0(\tau)$.
Subtracting those terms from \eq{gW-g}, nonsingular part of the gluon tensor is given by
%%%
\begin{align}
\label{eq:gW-gNS}
-g^{\mu\nu}\, W_{\mu\nu}^{g,\ns}
&= 4\as T_F Q_f^2(1-\e)
\bigg\{
\Theta_0(\tau,z)\big[
R^g(\tau,z)
+\delta(\tau-\tfrac{1-z}{z})
\Delta^g(z)
\big]
\nn\\
&\qquad
+(\Theta_0(\tau,z)-\theta(\tau)) P_{qg}(z) \frac{1}{\tau}
\bigg\}
\,.
\end{align}
%%%
Note that the last term in \eq{gW-gNS} gives  $1/\tau$ term in unphysical region where $\Theta_0$ is zero and this term cancel against the same term in singular part in the region.

Similarly, in the quark process $-g^{\mu\nu}W_{\mu\nu}^q$ in \eq{gW-g} contains terms with $\delta(\tau)$, $\cL_0(\tau)$, and $\cL_1(\tau)$ as $\tau\to 0$.
Subtracting those terms from \eq{gW-q}, the nonsingular part is given by
%%%
\begin{align}
-g^{\mu\nu}\, W^{q,\ns}_{\mu\nu}
&=2\as Q_f^2 C_F (1-\e)
\bigg\{
\frac{1}{\tau} \,N_{0}(\tau,z)
+\frac{\ln \tau}{\tau}\,N_{1}(\tau,z)
\nn\\
&\quad\quad
+\Theta_0(\tau,z)\,
\big[R^q(\tau,z)
+\delta\left(\tau-\frac{1-z}{z}\right)\,\Delta^q(\tau)
\big]\bigg\}
\label{eq:gW-qNS}
\,,
\end{align}
%%%
where the plus distribution $\cL_n$ is replaced by $\ln^n \tau/\tau$
and the function $N_{0,1}(\tau,z)$ is defined from differences as
%%%
\begin{align}
N_0(\tau,z)&=
\Theta_0(\tau,z)\, S_{0}^q(\tau,z)- \theta(\tau)S_{0}^q(0,z)
\nn\\
&=
2z\,[\Theta_0\,\cL_0(1-z,\tfrac{\tau}{1+\tau})- \theta(\tau)\cL_0(1-z)]
 -\frac{3}{2}\bigg[\Theta_0\,\deltaz 
\nn\\
&\qquad
- \theta(\tau)\delta(1-z)\bigg]
+(1-z)[\Theta_0(\tau,z)-\theta(\tau)]
\,,\nn\\ 
N_1(\tau,z)&=
\Theta_0(\tau,z)\,S_{1}^q(\tau,z)- \theta(\tau)S_{1}^q(0,z)
\nn\\
&=
-4\Big[\Theta_0\,\bigg(1+\frac{\tau}{2}\bigg)\, \deltaz
- \theta(\tau)\delta (1-z )\Big]
\label{eq:N01}
\,.
\end{align}
%%%

Now we calculate the cumulative of the nonsingular parts by using following integral 
%%%
\begin{align}\label{eq:Wcum}
\tW_{\mu\nu}(z,\tau,Q^2)=\int^\tau_0 d\tau'\,W_{\mu\nu}(z,\tau',Q^2)
\,.
\end{align}
%%%
Integration of \eqs{ppW-q}{ppW-g} is given by
%%%
\begin{align}
P^\mu P^\nu\, \tW^{q,\ns}_{\mu\nu}
&=\as C_F  Q_f^2 Q^2
\bigg[ \frac{1}{z}\min[\tau,\tfrac12,\tfrac{1-z}{z}]+\Theta_2 \frac{3z-2}{2z^2}
\bigg]
\,\nn\\
&=\as C_F  Q_f^2 Q^2
\left[\frac{\tau}{z}\, \Theta_0+\frac{1}{2z}(  \Theta_1 + \Theta_2 )\right]
\,,\nn\\
P^\mu P^\nu\, \tW^{g,\ns}_{\mu\nu}
&= 4\as T_F  Q_f^2 Q^2(1-z)
\left[\frac{\tau}{z}\, \Theta_0+\frac{1}{2z}(  \Theta_1 + \Theta_2 )\right]
\,,
\label{eq:ppWcum}
\end{align}
%%%
where $\Theta_0$ is defined in \eq{theta0} and 
%%%
\be 
\label{eq:Th12}
\Theta_{1}=\theta(-z+2/3)\,\theta(\tau-1/2)
\,,
\qquad\qquad
\Theta_2=\theta(z-2/3)\,\theta(\tau-\tfrac{1-z}{z})
\,.
\ee
%%%
They cover all regions in $\tau$ and $z$ space $\Theta_0+\Theta_1+\Theta_2=1$.
One finds that \eq{ppWcum} are discontinuous at $\tau=(1-z)/z$ when $z>2/3$
because $\delta(\tau-(1-z)/z)$ in \eqs{ppW-q}{ppW-g}
are turned on at the value and the discontinuity implies the end of physical region.

The cumulant of \eq{gW-gNS} is given by
%%%
\begin{align}
\label{eq:gWgcum}
-g^{\mu\nu}\, \tW_{\mu\nu}^{g,\ns}
&= 4\as T_F Q_f^2 (1-\e) \Bigl\{
\Theta_0\bigg[1-2\tau-P_{qg}(z)\ln\frac{1-\tau}{\tau}\bigg]-\big[1+P_{qg}(z)\ln\tau \big]
\Bigr\}
\,.
\end{align}
%%%
The cumulant of \eq{gW-qNS} is given by
%%%
\be\label{eq:gWqcum}
-g^{\mu\nu}\, \tW^{q,\ns}_{\mu\nu}
=2\as Q_f^2 C_F (1-\e)
\bigg\{
\tN_{0}(\tau,z) +\tN_{1}(\tau,z) +\tR^q(\taum,z)
+\Theta_2\Delta^q \Bigl(\frac{1-z}{z}\Bigr)
\bigg\}
\,,
\ee
%%%
where $\taum$ is the cumulative integral over $\Theta_0$
%%%
\be \label{eq:taum}
\taum\equiv \int d\tau' \Theta_0(\tau',z)
=\min \bigg\{\tau,\frac{1}{2},\frac{1-z}{z} \bigg\}
=\Theta_0\, \tau +\Theta_1 \, \frac{1}{2} + \Theta_2\, \frac{1-z}{z}
\,.
\ee
%%%

The functions $\tN_{0,1},\, \tR$ are defined as
%%%
\begin{subequations}
\begin{align}\label{eq:tNtR}
\begin{split}
\tN_n(\tau,z)&=\int_0^\tau d\tau'\, \frac{\ln^n \tau'}{\tau'} \,N_n(\tau',z)
\\
&=  \theta(\tau) \int_0^{\taum} \frac{d\tau'}{\tau'}\ln^n\tau' [S_{n}^q(\tau',z) -  S_n^q(0,z)] 
+ S_n^q(0,z) \Biggl[ 
 \frac{\ln^{n+1}\tau-\ln^{n+1}\taum}{n+1}
\Biggr]
\,,
\end{split}\\
\tR^q(\taum,z)&=\int^\tau_0 d\tau'\Theta_0(\tau',z) R^q(\tau',z)
=\frac{1-4z}{1-z}\taum-\frac{1+z^2}{1-z}\ln(1-\taum)
\,.\label{eq:Rqcumu}
\end{align}
\end{subequations}
%%%

%%%%%%%%%%%%%%%%%%%%%%%%%%%%%%%%%%%%%%%%%%%%%%%%%%%%%%%%%%%%%%%%%%%%%%%%%%%%%%%%
\subsection{Convolution with PDF} \label{ssec:convolution}

The hadronic tensor with a hadron state $h$ is expressed in the factorized form as
%%%
\begin{align}
W^h_{\mu\nu}(x,\tau,Q^2)
=\sum_{i\in \{q,\bar q,g\}} \int_x^1 \frac{d\xi}{\xi} f_{i/h}(\xi,\mu)\, w^{i}_{\mu \nu} (x/\xi,\tau,Q^2,\mu)
\label{eq:W-factor}
\,,
\end{align}
%%%
where $f_{i/h}$ is the PDF for the initial hadron $h$ into a parton $i$
and the superscript $i$ on the coefficient $w^i$ represents the
contribution from the parton $i$. Note that we obtained the tensors $W_{\mu\nu}^{q,g}$ for incoming quark and gluon and  by the perturbative matching performed in \cite{Kang:2014qba}
one can translate them into the coefficients $w^i$. We do not describe the matching procedure, and the expressions for $w^i$ are essentially the same as $W_{\mu\nu}^{q,g}$. 
Instead of writing $w^i$, we give expressions for the hadronic tensor \eq{W-factor} convolved with the PDF.  
The nonsingular part of the hadronic tensor is given by
%%%
\begin{align}\label{eq:ppW-proton}
&P^\mu P^\nu W^{h,\text{ns}}_{\mu\nu}
=\frac{ 2\pi Q^2}{x^2} (\cAns_q+\cAns_{\bar q} + \cAns_g)
\,,\\\nn
\\
\label{eq:cAq}
&
\cAns_{q} =
\sum_{f}Q_f^2 \frac{\as C_F}{4\pi}
\int_x^1 \frac{dz}{z} f_{q}\left(\frac{x}{z}\right)
\left[\Theta_0(\tau,z)\, 2z  +\delta(\tau-\tfrac{1-z}{z})\theta(z-2/3)\,(1-2\tau)(1+\tau)z^2)
\right]
\nn\\
&\quad \quad =
\sum_{f} Q_f^2 \frac{\as C_F}{4\pi}
\,
\Theta_0(\tau,x)
\left\{
\int_x^{\tfrac{1}{1+\tau}}dz\, 2\,f_{q}\bigg(\frac{x}{z}\bigg)
+\frac{1-2\tau}{(1+\tau)^2}f_{q}\big(x(1+\tau)\big)
\right\}
\,,\\
\label{eq:cAg}
&
\cAns_g =
\sum_f Q_f^2 \frac{\as T_F}{\pi}
\int_x^1 \frac{dz}{z} f_g\left(\frac{x}{z}\right) \Theta_0(\tau,z)
(1-z) \left[ 2z  +\delta(\tau-\tfrac{1-z}{z})\,(1-2\tau)(1+\tau)z^2
\right]
\nn\\
&\quad \quad
=
\sum_f Q_f^2 \frac{\as T_F}{\pi}
\,
\Theta_0(\tau,x)
\left\{
\int_x^{\tfrac{1}{1+\tau}}dz\, 2 (1-z) f_g\bigg(\frac{x}{z}\bigg)
+\frac{\tau(1-2\tau)}{(1+\tau)^3} f_g\big(x(1+\tau)\big)
\right\}
\,,
\end{align}
%%%

The sums goes over the flavors $f\in\{u,d,s,c,b\}$.
$\cAns_{\bar q}$ is obtained by replacing quark PDF  by anti-quark PDF in $\cAns_{q}$ and this is true for $\cBns_{q}$ and their cumulative.
%%%
The other tensor projection is expressed as
%%%
\be \label{eq:gW-proton}
-g^{\mu\nu} W_{\mu\nu}^{h,\text{ns}}= 8\pi(1-\e)(  \cBns_q + \cBns_{\bar q} + \cBns_g)
\,,
\ee
%%%
where $\cBns_q$ and $\cBns_g$ are
%%%
\begin{align}  \label{eq:cBq}
&
\cBns_{q} 
= \sum_{f} Q_f^2 \frac{\as C_F}{4\pi}
\bigg\{
\int_x^1\frac{dz}{z} f_{q,\bar q}\left(\frac{x}{z}\right)
\bigg[ \frac{1}{\tau} \,N_{0}(\tau,z)+\frac{\ln \tau}{\tau}\,N_{1}(\tau,z)
\bigg]
\nn\\
&\qquad\qquad
+\Theta_0(\tau,x)
\int_x^{\tfrac{1}{1+\tau}}\frac{dz}{z} f_{q,\bar q}\left(\frac{x}{z}\right) R^q(\tau,z)
+\Theta_0(\tau,x) \,f_{q,\bar q}(x(1+\tau))\frac{\Delta^q(\tau)}{1+\tau}
\bigg\}
\,,\\
 \label{eq:cBg}
& 
\cBns_g =
 \sum_f Q_f^2 \frac{\as T_F}{2\pi}
\bigg\{
\Theta_0 \int_x^{\tfrac{1}{1+\tau}}\frac{dz}{z} f_g\big(x/z\big)
\bigg(\frac{P_{qg}(z)}{1-\tau}-2\bigg)
\nn\\&\hspace{3cm}
+\Theta_0  f_g(x(1+\tau))
\bigg(\frac{2\tau-1}{1+\tau}+\frac{1+\tau^2}{(1+\tau)^3}\ln\frac{1-\tau}{\tau}  \bigg)
\nn\\&\hspace{3cm}
+\frac{1}{\tau}\Theta_0\int^{\onet}_x \frac{dz}{z}f_g\left(\frac{x}{z}\right)P_{qg}(z)-\frac{1}{\tau}\int^1_x \frac{dz}{z}f_g\left(\frac{x}{z}\right)P_{qg}(z)
\bigg\}
\,, 
\end{align}
%%%
where the integrals involving $N_{0,1}$ are given by
%%%
\begin{align}\label{eq:intN01}
&\int_x^1\frac{dz}{z} f_q\left(\frac{x}{z}\right)
\frac{\ln \tau}{\tau} \,N_1 (\tau,z)
=-4\frac{\ln\tau}{\tau}\bigg[\Theta_0\,
\frac{2+\tau}{2(1+\tau)}f_q\big(x(1+\tau)\big)
- f_q(x)
\bigg]
\,,\\
&\int_x^1\frac{dz}{z} f_q\left(\frac{x}{z}\right)
\frac{1}{\tau} \,N_{0}(\tau,z)
=-\frac{3}{2\tau} \bigg[\Theta_0\,\frac{f_q\big(x(1+\tau)\big)}{1+\tau}-f_q(x)
\bigg]
\nn\\
&\quad
+\frac{2}{\tau} \bigg\{
\Theta_0\bigg[ \int^{\onet}_{x} dz \frac{f_q\big(x/z\big)-f_q\big(x(1+\tau)\big)}{1-z}+f_q\big(x(1+\tau)\big)\ln(1-x)
\bigg]
\nn\\
&\hspace{2cm}
-\int^1_x dz\frac{f_q\big(x/z\big)-f_q(x)}{1-z}-f_q(x)\ln(1-x)
\bigg\}
\nn\\&\quad
+\frac{1}{\tau} \bigg\{
\Theta_0\,
\int^{\tfrac{1}{1+\tau}}_x dz\,f_q\left(\frac{x}{z}\right)\frac{1-z}{z}
-\int_x^1dz\,f_q\left(\frac{x}{z}\right)\frac{1-z}{z}
\bigg\}
\,.
\end{align}
%%%
Note that terms with $1-\Theta_0=\Theta_1+\Theta_2$ are contributions from unphysical regions and they cancel the contributions from singular parts in the same region.

Next let us calculate the convolution with cumulants of \eq{ppWcum}
%%%
\begin{align}
\label{eq:pptW-proton}
P^\mu P^\nu \tW^{h,\text{ns}}_{\mu\nu}
&=\frac{ 2\pi Q^2 }{x^2} (\Ans_{q}+ \Ans_{\bar q} + \Ans_{g})
\,,\\\nn\\
\label{eq:Aq}
\Ans_{q}&=
\sum_{f} Q_f^2 \frac{\as C_F}{4\pi}  \int_x^1 \frac{dz}{z} f_{q,\bar q}\left(\frac{x}{z}\right) \,\bigg[ 2z\taum + \Theta_2 \, (3z-2)  \bigg]
\nn\\&=
\sum_{f} Q_f^2 \frac{\as C_F}{4\pi}
\,\bigg\{
2\tau \Theta_0\,\int^{\tfrac{1}{1+\tau}}_x dz\,f_q\left(\frac{x}{z}\right)+\Theta_1\int^{\tfrac{2}{3}}_x dz\,f_q\left(\frac{x}{z}\right)
\nn\\& \qquad \hspace{6cm}
+\int^1_{\max\big[\tfrac{1}{1+\tau},\tfrac{2}{3},x\big]}dz\,f_q\left(\frac{x}{z}\right)
\bigg\} 
\nn \\
& = \sum_f  Q_f^2 \frac{\as C_F}{4\pi}\biggl\{(2\tau-1)\, \Theta_0 \int_x^{\frac{1}{1+\tau}} dz \, f_{q} \left(\frac{x}{z}\right)  + \int_x^1 dz\, f_{q}\left(\frac{x}{z}\right)\biggr\}
\,,\\
\label{eq:Ag}
\Ans_{g}
 &= \sum_f Q_f^2 \frac{\as T_F}{\pi} \biggl\{(2\tau-1)\, \Theta_0 \int_x^{\frac{1}{1+\tau}} dz f_g\left(\frac{x}{z}\right)  (1-z)  + \int_x^1 dz\, f_g\left(\frac{x}{z}\right) (1-z)\biggr\}
\,.
\end{align}
%%%
Here the extra factor $(1-z)$ in the gluon integral compared with the quark part corresponds to the extra factor $(1-z)$ in \eq{ppWcum}.

Now we calculate convolution with cumulants in \eqs{gWgcum}{gWqcum}:
%%%
\be\label{eq:gtW-proton}
-g^{\mu \nu} \tW^{h,\text{ns}}_{\mu\nu}
=8 \pi (1-\e) (\Bns_{q} + \Bns_{\bar q} + \Bns_{g})
\,,
\ee
%%%
where
%%%
\begin{align}
\label{eq:Bg}
\Bns_{g} &=
 \sum_f Q_f^2 \frac{\as T_F}{2\pi} \biggl\{
 \Theta_0 \int^{\onet}_{x}\frac{dz}{z}\,f_g\left(\frac{x}{z}\right)
 \big[1-2\tau - P_{qg}(z)\,\ln\tfrac{1-\tau}{\tau}\big]
\nn\\ 
& \qquad \qquad \qquad \qquad 
 -\int^1_x \frac{dz}{z}\, f_g\left(\frac{x}{z}\right)\big[1+P_{qg}(z)\ln\tau  \big]
  \biggr\}
\,,\\
\label{eq:Bq}
\Bns_{q}&=
\sum_{f} Q_f^2 \frac{\as C_F}{4\pi}
\bigg\{
\int_0^\tau d\tau'\int_x^1 \frac{dz}{z} f_{q}\left(\frac{x}{z}\right) \,
\bigg[ \frac{1}{\tau} N_{0}(\tau',z) +\frac{\ln \tau'}{\tau'}N_{1}(\tau',z)
\bigg]
\nn\\&\hspace{3cm}
+\,\int^{1}_x \frac{dz}{z} f_{q}\left(\frac{x}{z}\right)\, \widetilde{R}^q(\tau',z)+ \,\int^{1}_x \frac{dz}{z} f_{q}\left(\frac{x}{z}\right) \widetilde{\Delta}^q(\tau,z)
\bigg\}
\,,
\end{align}
%%%
where $\tR^q$ is cumulant of $R^q$ and is defined in \eq{Rqcumu} and $\widetilde{\Delta}^q$ is cumulant of $\delta^q$ term, that is
%%%
\be\label{eq:delta-q-cumu}
\widetilde{\Delta}^q (\tau,z)=\Theta_2 (\tau,z)\Delta^q\big((1-z)/z \big).
\ee
%%%
Plugging \eqs{Rqcumu}{delta-q-cumu} into last two integrals in \eq{Bq}, we obtain
%%%
\begin{align}\label{eq:RqDelta}
 &\int^{1}_x \frac{dz}{z} f_{q}\left(\frac{x}{z}\right)\, \bigg(\widetilde{R}^q(\tau,z)+ \widetilde{\Delta}^q(\tau,z)\bigg)
 \nn \\& \quad
  =\Theta_0\int^{\onet}_x \frac{dz}{z} f_q\left(\frac{x}{z}\right)\bigg[
 \frac{1-4z}{1-z}\tau-P_{qq}(z)\,\ln (1-\tau)
 \bigg]
 \nn\\
 &\qquad
 +\Theta_1 \int^{\frac23}_{x}\frac{dz}{z}\,f_q\left(\frac{x}{z}\right)\bigg[
 \frac{1-4z}{2(1-z)}+P_{qq}(z)\,\ln 2 \bigg]
 \nn \\& \quad\quad
 +\int^1_{\max\big[\onet,\frac{2}{3},x\big]}\frac{dz}{z}f_q \left(\frac{x}{z}\right) \bigg[
 \frac12+\frac{2\ln z}{z-1}+(z-1)\ln \frac{1-z}{z}
 \bigg]
 \,.
\end{align}
%%%
The terms $N_i$ are given in \eq{N01}.
Integral of each term is given by
%%%
\begin{align}
&\int_0^1 d\tau'\int_x^1 \frac{dz}{z} f_q\left(\frac{x}{z}\right) \,\frac{\ln \tau'}{\tau'} N_1(\tau',z)
\nn\\&\quad=
-4\bigg[
\int_0^{\taum} d\tau'\,\frac{\ln \tau'}{\tau'}\bigg(\frac{2+\tau'}{2(1+\tau')}f_q(x(1+\tau'))-f_q(x)\bigg)
- \frac{f_q(x)}{2} \left( \ln^2\tau-\ln^2 \taum]  \right)
\bigg]
\nn\\
&\quad=
-2\int^1_{\max\big[\onet,\frac23,x\big]}\frac{dz}{z}\frac{\ln\frac{1-z}{z}}{1-z}
\bigg[
(1+z)f_q\left(\frac{x}{z}\right)-2f_q(x)
\bigg]
+2f_q(x)\,\bigg[\ln^2 \tau - \ln^2\taum \bigg]
\,,\label{eq:intN1}
\\ \nn
\\&
\int_0^\tau d\tau'\int_x^1 \frac{dz}{z} f_q\left(\frac{x}{z}\right) \,\frac{1}{\tau'}
 N_0(\tau',z)
\nn\\&\quad=
2\!\!
\int_0^{\taum} d\tau'\,
\bigg[
\frac{\ln\tfrac{ \tau'}{1+\tau'}}{\tau'}\big[f_q(x(1+\tau'))-f_q(x)\big]
-\frac{1}{\tau'}\!\!\int^1_{\tfrac{1}{1+\tau'}}\, dz\frac{f_q\left(\frac{x}{z}\right)-f_q(x)}{1-z}
\bigg]
\nn\\&\hspace{1cm}
-2\ln\bigg(\frac{\tau}{\taum}\bigg)
\bigg(\int_x^1 dz\, \frac{f_q\left(\frac{x}{z}\right)-f_q(x)}{1-z}+f_q(x) \ln(1-x) \bigg)
\nn\\
&\hspace{1cm}
-\frac32 \bigg[
\int_0^{\taum} d\tau'\,\frac{1}{\tau'}\bigg[\frac{1}{(1+\tau')}f_q(x(1+\tau'))-f_q(x)\bigg]
- f_q(x) \left( \ln\tau-\ln\taum \right)
\bigg]
\nn\\
&\hspace{1cm}+
\int_x^1 dz\, \frac{(1-z)\,f_q\left(\frac{x}{z}\right)}{z} \ln \frac{\taum }{\tau}
\label{eq:intN0}
\,.
\end{align}
%%%
The double integrals over $\tau'$ and $z$ that still exist in \eq{intN0} can be simplified by switching the order of integration. Doing so, and also changing variables in the remaining integrals over $\tau'$ using $z = 1/(1+\tau')$ so that all integrals are over $z$, we obtain simpler expression
%%%
\begin{align} \label{eq:N0simpler}
&\nn
\int_0^\tau d\tau'\int_x^1 \frac{dz}{z} f_q\left(\frac{x}{z}\right) \,\frac{1}{\tau'} N_0(\tau',z) 
\nn\\
&=
 2 \intmax
 \frac{dz}{z}\Bigl[ \ln (1-z) + z\ln\frac{1-z}{z\tau}\Bigr]\frac{f_q\left(\frac{x}{z}\right)-f_q(x)}{1-z} 
  \\
& \quad
 - 2 \Theta_1 \ln(2\tau) \int_x^{\frac{2}{3}} dz\frac{f_q\left(\frac{x}{z}\right)-f_q(x)}{1-z} - 2\Bigl[\Theta_1 \ln(2\tau) + \Theta_2\ln\frac{x\tau}{1-x}\Bigr] f_q(x)\ln(1-x) 
 \nn \\
& \quad 
- \frac{3}{2}\intmax \frac{dz}{z} \frac{ zf_q\left(\frac{x}{z}\right) - f_q(x)}{1-z} + \frac{3}{2} f_q(x) \Bigl[ \Theta_1 \ln(2\tau) + \Theta_2 \ln \frac{\tau x}{1-x}\Bigr] 
\nn \\
&\quad
 + \intmax dz\,f_q\left(\frac{x}{z}\right)\frac{1-z}{z} \ln\frac{1-z}{z\tau} - \Theta_1 \ln(2\tau) \int_x^{2/3} dz \,f_q\left(\frac{x}{z}\right)\frac{1-z}{z}\,.
\end{align}
%%%
Finally, we combine all terms contributing to $\Bns_q$ in \eq{Bq}. Some further simplifications occur upon summing \eqss{RqDelta}{intN1}{N0simpler}. Doing so, we obtain the final result for $\Bns_q$, which can be written as in \eq{AnsBns}.

%%%%%%%%%%%%%%%%%%%%%%%%%%%%%%%%%%%%%%%%%%%%%%%%%%%%%%%%%%%%%%%%%%%%%%%%%%%%%%%%%%%%%%
\section{Additional contribution for $\tauct$}\label{app:tauct}
Here, we summarize the results of additional contributions for $\tauct$ which 
is to be added onto $\taujt$ results. 

Since we take the difference between two 1-jettiness, all divergences  are cancelled. Hence, the necessary $v$ integrals are finite and all $\e$ can be dropped as done in \eq{dPhi2-IV} and in the quark amplitudes \eqs{gMM-real}{ppM-real} and in the gluon amplitudes \eqs{gMM-g}{ppM-g}.
Integrating over $v$ in these amplitudes and performing the matching, we obtain the final results.
To avoid duplication we omit showing intermediate steps and actually the intermediate expressions are similar to final results because there are not much simplifications.

The final expressions of differential version of $\delta A$ and $\delta B$ are given by
%%%
\begin{align}\label{eq:delcAcB}
    \delta \cAns_q &= \sum_f Q^2_f\frac{\as C_F}{4\pi}\,
    \left\{
    \frac{2z_0^4}{3z_0-2} \,  \delta \hat A_q(R,z_0) 
   -  2z_1^3\,  \delta \hat A_q(R,z_1)
    \right\}
  \,,   
    \nn\\
    \delta \cAns_g &= \sum_f Q^2_f\frac{\as T_F}{\pi}\,
        \left\{
     \frac{z_0^4}{3z_0-2} \,   \delta \hat A_g(R,z_0)
   -  z_1^3\,  \delta \hat A_g(R,z_1)
    \right\}
    \,,
    \nn\\
    \delta \cBns_q &= \sum_f Q^2_f\frac{\as C_F}{4\pi}\,
       \left\{
     \frac{z_0^2}{3z_0-2} \,   \delta \hat B_q(R,z_0)
   -  z_1\,  \delta \hat B_q(R,z_1)
    \right\}
    \,,
    \nn\\
    \delta \cBns_g &= \sum_f Q^2_f\frac{\as T_F}{2\pi}\,
    \left\{
     \frac{z_0^2}{2(3z_0-2)} \,   \delta \hat B_g(R,z_0)
   - \frac{z_1}{2}\,  \delta \hat B_g(R,z_1)
       \right\}
       \,.
\end{align}
%%%
The parameters $z_{0,1}$ are the relations between $z$ and $\tau$ 
in region IV
%%%
\begin{align}\label{eq:z0z1}
z_0=\frac{3+\sqrt{1-4\tau}}{2(2+\tau)}\,,\quad
z_1=\frac{1}{1+\tau}\,.
\end{align}
%%%
The functions $\delta \hat A,\delta \hat B$ are
%%%
\begin{align}\label{eq:hatAB}
\delta \hat A_q(R,z)&=
   \theta(z-x)\theta(z-z_c)\theta(-z+1)
    \frac{r(z,R)}{2z}
    f_q\left(\frac{x}{z}\right)\,,
\nn\\
\delta \hat A_g(R,z) &=
\theta(z-x)\theta(z-z_c)\theta(-z+1)
\frac{1-z}{z}r(z,R) f_g\left(\frac{x}{z}\right)
 \,,\nn\\
\delta \hat B_q(R,z) &=
 \theta(z-x)\theta(z-z_c)\theta(-z+1)
 \left[
    \frac{1-4z}{2(1-z)}r(z,R)+\frac{1+z^2}{1-z}\ln\frac{1+r(z,R)}{1-r(z,R)}
   \right] 
   f_q\left(\frac{x}{z}\right)\,,\nn\\
\delta \hat B_g(R,z) &=
     \theta(z-x)\theta(z-z_c)\theta(-z+1)
     \left[
     2P_{qg}(z) \ln\frac{1+r(z,R)}{1-r(z,R)} -2r(z,R)
   \right]
   f_g\left(\frac{x}{z}\right)
   \,,
\end{align}
%%%
where $z_c(R)$ is given in \eq{vpm-zc} and $r(z,R)$ is in \eq{r}.
If we insert $z_0$ and $z_1$ into the constraint $\theta(z-z_c)$, we obtain two upper limits of $\tau$ $(16-R^2)R^2/256$ and $R^2/16$, beyond which each of two contributions in \eq{delcAcB} becomes zero.

%%%%%%%%%%%%%%%%%%%%%%%%%%%%%%%%%%%%%%%%%%%%%
\section{Difference between $\taub$ and $\taujt$}  \label{app:diff}

Here we give the difference between $\taub$ and  $\taujt$ results.
The cumulative results for $\taujt$ are given in \eqs{AsBs}{AnsBns} and results for $\taub$ in \cite{Kang:2014qba}.
Their differences in singular and nonsingular parts are given by 
%%%
\begin{align}\label{eq:taub-taujt}
 B_q^\sing(\taub)- B_q^\sing (\taujt)&=
  \sum_f  Q_f^2\frac{\as C_F}{4\pi} \int _x^1 \frac{dz}{z} f_q\left(\frac{x}{z}\right) P_{qq}(z)\ln z
 \,, \nn\\
 B_g^\sing(\taub)- B_g^\sing (\taujt)&=
  \sum_f Q_f^2 \frac{\as T_F}{2\pi} 
\int_x^1 \frac{dz}{z}f_g\left(\frac{x}{z}\right) P_{qg}(z)\ln z
 \,, \nn\\
A_q^\ns(\taub)- A_q^\ns (\taujt)&=
\sum_f  Q_f^2 \frac{\as C_F}{4\pi}\, \Theta_0 \int_x^{\frac{1}{1+\tau}} dz \, f_{q} \left(\frac{x}{z}\right)  2\tau(z-1)
\,, \nn\\
A_g^ \ns(\taub)- A_g^\ns (\taujt)&=
\sum_f Q_f^2 \frac{\as T_F}{\pi} \, \Theta_0 \int_x^{\frac{1}{1+\tau}} dz f_g\left(\frac{x}{z}\right)  2\tau[-(1-z)^2]
\,,  \nn\\
B_q^\ns(\taub)- B_q^\ns (\taujt)&=
\sum_f Q_f^2 \frac{\as C_F}{4\pi}
 \Biggl( \Theta_0
\biggl\{
\int_{x}^{\frac{1}{1+\tau}} \frac{dz}{z} f_q\left(\frac{x}{z}\right) \Bigl[-\tau(1-4z) + P_{qq}(z)\ln\frac{z(1-\tau)}{1-z\tau}\Bigr]\biggr\}
\nn \\ & \quad
+\int^1_x \frac{dz}{z}\,f_q\left(\frac{x}{z}\right)\bigg[-P_{qq}(z)\ln z \bigg]
\Biggr)
\,, \nn\\
B_g^ \ns(\taub)- B_g^\ns (\taujt)&=
\sum_f Q_f^2 \frac{\as T_F}{2\pi} \biggl\{
 \Theta_0 \int^{\onet}_{x}\frac{dz}{z}\,f_g\left(\frac{x}{z}\right)
 \bigg[2\tau(1- z)+ P_{qg}(z)\,\ln\frac{z(1-\tau)}{1-z\tau}\bigg]
 \nn\\
 &\qquad\qquad
 -\int^1_x \frac{dz}{z}\, f_g\left(\frac{x}{z}\right)
 P_{qg}(z)\ln z 
  \biggr\}
\,,
\end{align}
%%%
where those not listed above are zero.

The NLP obtained from \eq{taub-taujt}  can be expressed as 
\begin{align}\label{eq:taub-taujtex}
A_q^\ns(\taub)- A_q^\ns (\taujt)\big|_{\tau\rightarrow 0}&=
\sum_f  Q_f^2 \frac{\as C_F}{4\pi}\,  \tau\int_x^{1} dz \, f_{q} \left(\frac{x}{z}\right)  2(z-1)
+\cO(\tau^2)\,, \nn\\
A_g^ \ns(\taub)- A_g^\ns (\taujt)\big|_{\tau\rightarrow 0}&=
\sum_f Q_f^2 \frac{\as T_F}{\pi} \,  \tau\int_x^{1} dz f_g\left(\frac{x}{z}\right)  [-2(1-z)^2]
+\cO(\tau^2)\,,  \nn\\
B_q^\ns(\taub)- B_q^\ns (\taujt)\big|_{\tau\rightarrow 0}&=
\sum_f Q_f^2 \frac{\as C_F}{4\pi}\tau\biggr\{
\int_x^{1} \frac{dz}{z}\, f_q\left(\frac{x}{z}\right) \bigg[-1+4z+P_{qq}(z)(z-1)\bigg]
\nn\\ 
&\qquad\qquad \qquad\qquad 
+2f_q\left(x\right)\biggr\}
+\cO(\tau^2)\,, \nn\\
B_g^ \ns(\taub)- B_g^\ns (\taujt)\big|_{\tau\rightarrow 0}&=
\sum_f Q_f^2 \frac{\as T_F}{2\pi} \tau\biggr\{
\int_x^{1}\frac{dz}{z}\, f_g\left(\frac{x}{z}\right) 
(1-z)\big[2-P_{qg}(z)\big]\biggr\}
+\cO(\tau^2)\,.
\end{align}
%%%%%%%%%%%%%%%%%%%%%%%%%%%%%%%%%%%%%%%%%%%%%
\bibliography{paper3}
\end{document}